# The accomplishment of the Engineering Design Activities of IFMIF/EVEDA: The European–Japanese project towards a Li(d,xn) fusion relevant neutron source

**J. Knaster[1], A. Ibarra[2], J. Abal[3], A. Abou-Sena[4], F. Arbeiter[4], F. Arranz[2], J.M. Arroyo[2], E. Bargallo[3], P-Y. Beauvais[5], D. Bernardi[6], N. Casal[2], J.M. Carmona[2], N. Chauvin[5], M. Comunian[11], O. Delferriere[5], A. Delgado[2], P. Diaz-Arocas[2], U. Fischer[4], M. Frisoni[6], A. Garcia[2], P. Garin[1], R. Gobin[5], P. Gouat[7], F. Groeschel[1], R. Heidinger[8], M. Ida[9], K. Kondo[10], T. Kikuchi[9], T. Kubo[1], Y. Le Tonqueze[1], W. Leysen[7], A. Mas[2], V. Massaut[7], H. Matsumoto[1], G. Micciche[6], M. Mittwollen[4], J.C. Mora[2], F. Mota[2], P.A.P. Nghiem[5], F. Nitti[8], K. Nishiyama[1], F. Ogando[12], S. O'hira[10], C. Oliver[2], F. Orsini[5], D. Perez[2], M. Perez[1], T. Pinna[6], A. Pisent[11], I. Podadera[2], M. Porfiri[6], G. Pruneri[1], V. Queral[2], D. Rapisarda[2], R. Roman[2], M. Shingala[1], M. Soldaini[1], M. Sugimoto[10], J. Theile[13], K. Tian[4], H. Umeno[1], D. Uriot[5], E. Wakai[14], K. Watanabe[14], M. Weber[2], M. Yamamoto[1] and T. Yokomine[15]**

[1] IFMIF/EVEDA Project Team, Rokkasho, Japan
[2] CIEMAT, Madrid, Spain
[3] UPC (Universidad Politécnica de Cataluña), Spain
[4] KIT, Karlsruhe, Germany
[5] CEA, Saclay, France
[6] ENEA, Brasimone Italy
[7] SCK-CEN, Mol, Belgium
[8] F4E, Garching, Germany
[9] JAEA, Oarai, Japan
[10] JAEA, Rokkasho, Japan
[11] INFN, Legnaro, Italy
[12] UNED, Madrid, Spain
[13] CRPP, Villigen, Switzerland
[14] JAEA, Tokai, Japan
[15] Kyoto University, Japan

E-mail: juan.knaster@ifmif.org

### Abstract

The International Fusion Materials Irradiation Facility (IFMIF), presently in its Engineering Validation and Engineering Design Activities (EVEDA) phase under the frame of the Broader Approach Agreement between Europe and Japan, accomplished in summer 2013, on schedule, its EDA phase with the release of the engineering design report of the IFMIF plant, which is here described. Many improvements of the design from former phases are implemented, particularly a reduction of beam losses and operational costs thanks to the superconducting accelerator concept, the re-location of the quench tank outside the test cell (TC) with a reduction of tritium inventory and a simplification on its replacement in case of failure, the separation of the irradiation modules from the shielding block gaining irradiation flexibility and enhancement of the remote handling equipment reliability and cost reduction, and the water cooling of the liner and biological shielding of the TC, enhancing the efficiency and economy of the related sub-systems. In addition, the maintenance strategy has been modified to allow a shorter yearly stop of the irradiation operations and a more careful management of the irradiated samples. The design of the IFMIF plant is intimately linked with the EVA phase carried out since the entry into force of IFMIF/EVEDA in June 2007. These last activities and their on-going accomplishment have been thoroughly described elsewhere (Knaster J *et al* [19]), which, combined with the present paper, allows a clear understanding of the maturity of the European–Japanese international efforts. This released IFMIF Intermediate Engineering Design Report (IIEDR), which could be complemented if required concurrently with the outcome of the on-going EVA, will allow decision making on its construction and/or serve as the basis for the definition of the next step, aligned with the evolving needs of our fusion community.

## 1. Introduction

The safe design of a fusion power reactor is essential for getting the license for operation granted by the corresponding Nuclear Regulatory Agency. As relevant as confining the plasma in a stable manner under fusion conditions is the use of suitable materials for the in-vessel components, capable of withstanding the severe operational conditions without being degraded either in their dimensional stability, or in their mechanical and physical properties beyond allowable design levels. Furthermore, alloying elements either forming long-lived radioisotopes or causing substantial decay heat have to be reduced to a minimum level.

The complexity of the radiation damage mechanisms in materials, due to a superposition of transmutation products, displacement damage, thermo-mechanical loads and corrosion/erosion enhancement, calls for experimental studies under conditions as close as possible to realistic cases. The diversity of key parameters involved (neutron flux, spectrum, fluence, material temperature, mechanical loading conditions, microstructure, thermo-mechanical processing history, lattice kinetics etc) makes existing material models incomplete. Thus, a neutron source with a suitable fluence and spectrum becomes an unavoidable step in the design and construction of fusion reactors subsequent to ITER, where potential structural damage rates exceeding 15 $dpa_{NRT}$ (displacement per atom/ Norgett, Robinson and Torrens) [1] (meaning $dpa_{NRT}$ iron equivalent) per year of operation are expected for a 3.0 GW thermal power DEMO concept reactor [2] compared with less than 3 $dpa_{NRT}$ for ITER at the end of its designed life.

The seminal proposal towards a fusion relevant neutron source based on Li(d,xn) nuclear reactions was published in 1976 [3]. As early as 1979, the first review of the state-of-the-art underlying technology concluded that such a neutron source was indispensable [4].

## 2. The genealogy of the International Fusion Materials Irradiation Facility

Different concepts have been in place since the 1970s, when the Fusion Materials Irradiation Test project (FMIT) was proposed in the United States of America (USA) [5]. FMIT aimed at obtaining a neutron flux of $10^{19}$ $m^{-2}$ $s^{-1}$ in a 10 $cm^3$ volume with a 35 MeV deuteron accelerator of 100 mA current in CW (continuous wave, meaning 100% duty cycle) colliding on a flowing lithium screen (see figure 1).

However, the need for a fusion relevant neutron source, together with the technological endeavours of learning how to confine the plasma, was not so apparent at the time without fusion power on the horizon, and the project was stopped in 1984 although positive results of the validation activities had been obtained [6]. The International Energy Agency (IEA) fostered a series of regional meetings (in the USA, Europe, and Japan) throughout 1988, which culminated in early 1989 in an international workshop to select the most promising candidate [7] for a fusion neutron source. Consensus was attained within the material scientist community that an accelerator-based neutron source utilising Li(d,xn) nuclear stripping reactions [8] would be the optimal choice. Aligned with this, JAERI's timely proposal of the Energy Selective Neutron Irradiation Test Facility project (1988–92) with 50 mA CW, 40 MeV deuteron beam and a 125 $cm^3$ testing volume with a neutron flux of $3\times10^{18}$ $m^{-2}$ $s^{-1}$ [9, 10], coincided with parallel, but less successful, initiatives in the USA [11].

Through international advisory boards coordinated by the IEA, a neutron source based on Li(d,xn) nuclear reactions was acclaimed in 1992 [12]. Since 1994, the International Fusion Materials Irradiation Facility (IFMIF) is the reference concept within the fusion community. The design baseline was documented in the final report of its Conceptual Design Activity (CDA) phase issued in 1996 [13] as the outcome of a joint effort of the European Union, Japan, the Russian Federation, and USA within the framework of the Fusion Materials Implementing Agreement of the IEA. A cost estimate [14] for IFMIF was developed during the CDA phase, which entailed further design studies in 1997 and 1998 resulting in the Conceptual Design Evaluation report [15]. In 1999, the IEA Fusion Power Coordinating Committee asked for a review of the IFMIF design, and stipulated to focus on cost reduction [16] while safeguarding the original mission. The 'key element technology phase' implemented those directives during 2000–2002 with the objectives of: (1) reducing the key technology risk factors on the way to achieve a CW deuteron beam with the needed current; (2) verifying relevant component designs on a laboratory scale (both in the lithium target system and test facilities (TFs)); and (3) validating design codes [17].

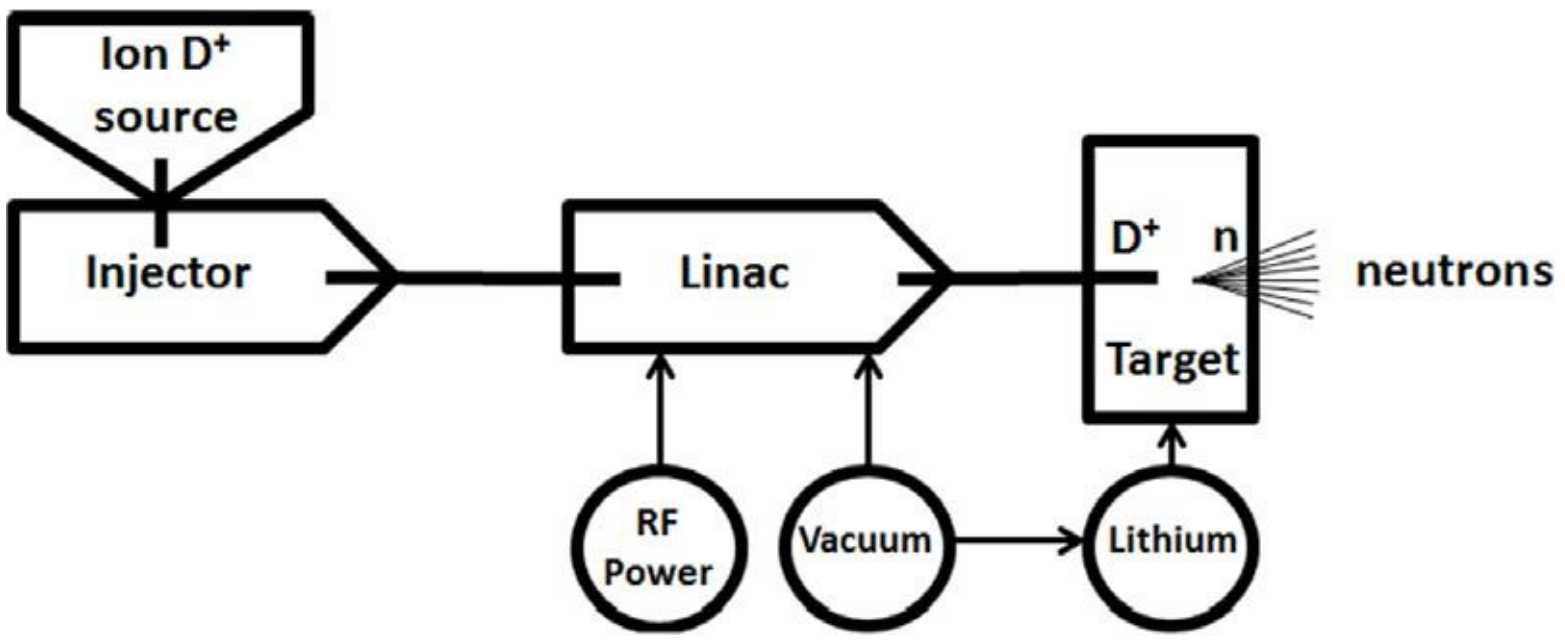

**Figure 1.** Principle of a Li(d,xn) neutron source as proposed for FMIT (as per figure 1 of [5]).

In 2004, the Conceptual Design Report co-authored by a team from the four aforementioned countries was released [18]. World-wide discussions preceded the approval of the IFMIF/EVEDA (Engineering Validation and Engineering Design Activities) project in 2007, concurrently with the ITER Agreement. The IFMIF/EVEDA project is one of three projects defined in the Broader Approach agreement between Japan and EURATOM, which entered into force in June 2007. The IFMIF/EVEDA specific Annex in the BA Agreement mandates the project to produce an integrated engineering design of IFMIF and the data necessary for future decisions on the construction, operation, exploitation and decommissioning of IFMIF, and to validate the continuous and stable operation of each IFMIF subsystem. Though the validation activities were not fully completed when the Engineering Design Activities (EDA) phase ended in June 2013, within the six years allocated, maturity of the studies allowed the successful development of the IFMIF Intermediate Engineering Design Report (IIEDR) that is here described. In turn, the status of the project and of the Engineering Validation Activities (EVA) phase at the time of the accomplishment of the EDA phase has been reported elsewhere [19–21].

## 3. The scope of IFMIF in the EVEDA phase

The continuous worldwide activities carried out since the initial proposal in 1976 of a Li(d,xn) fusion relevant neutron source [3] produced a solid conceptual design of IFMIF in 2004 [18], that was the starting point for the definitive EVEDA phase as an effective risk mitigation exercise before facing its construction. Clearly, notwithstanding the technological interest of a Li(d,xn) facility, which involves attractive high current accelerators and liquid alkaline metals and cutting-edge technologies, the materialisation of IFMIF is directly associated with the ITER and DEMO schedule. A key factor for the construction of the latter is the timely availability of a conclusive database for structural materials including their radiation-induced degradation. Validated data are essential for the lifetime evaluation, final design, licensing and reliable operation of DEMO components; data from IFMIF are to be obtained within the same time frame as results from the ITER operation.

To ensure that the needs in materials science and fusion technology were fulfilled at this final stage before construction, a specification working group was formed soon after the start of the EVEDA phase [22] in order to update the IFMIF users requirements defined in 1998 [23]. The major points of this specification working group can be summarised as follows: (1) irradiation programme for structural materials will consist of three exposure types to focus on data taking up to 50, 100 and 150 dpa levels; (2) damage production in the high flux region shall be >20 dpa/fpy (full power year) in 0.5 l volume (with neutron flux gradient and irradiation temperature variations inside the gauge volume of the set of samples within 10% and $\pm$3%); (3) the design lifetime of the plant is 30 years to cover all planned irradiation tests; (4) use of small specimens is considered for both the post-irradiation examination facility (PIEF) and *in-situ* type experiments; and (5) some tests of non-structural materials will also be possible in the high flux region.

The main expected IFMIF contributions are to: (1) provide data for the engineering design for DEMO; (2) provide information to define performance limits of materials and material systems for DEMO and beyond; (3) contribute to the completion and validation of (existing) databases to collect and confirm data required for licensing and safety assessment; (4) contribute to the selection or optimisation of different alternative fusion materials; (5) validate the fundamental understanding of the radiation response of materials including the benchmarking of irradiation effects modelling at a length-scale and time-scale relevant for engineering application; and (6) test characteristic components of blankets prior to or complementary to ITER test blanket modules [22, 23].

## 4. The engineering design of IFMIF

IFMIF will generate a neutron flux with a broad peak at 14 MeV by Li(d,xn) nuclear reactions thanks to two parallel deuteron beams colliding onto a liquid lithium jet. The two accelerators will generate beams of 40 MeV and currents of 125 mA each in CW mode with a common footprint of 200 mm $\times$ 50 mm. The beam energy has been tuned to obtain a neutron spectrum which simulates the best irradiation effects occurring in the first wall of a fusion reactor. The flux area extends over a volume of 500 $cm^3$ for which a test module (TM) is developed that can accommodate around 1000 quasi-standardised specimens. The design is made to allow for 30 years of operation [22].

The IFMIF plant is composed of five specific facilities. Accordingly, the systems designed for the IFMIF plant are grouped into the accelerator facility (AF), the lithium target facility (LF), the TF, PIEF and, the conventional facilities (CFs). The latter group of systems ensure power, cooling, ventilation, rooms and services to the other facilities and itself [21]. A schematic of the systems of the IFMIF plant is shown in figure 2.

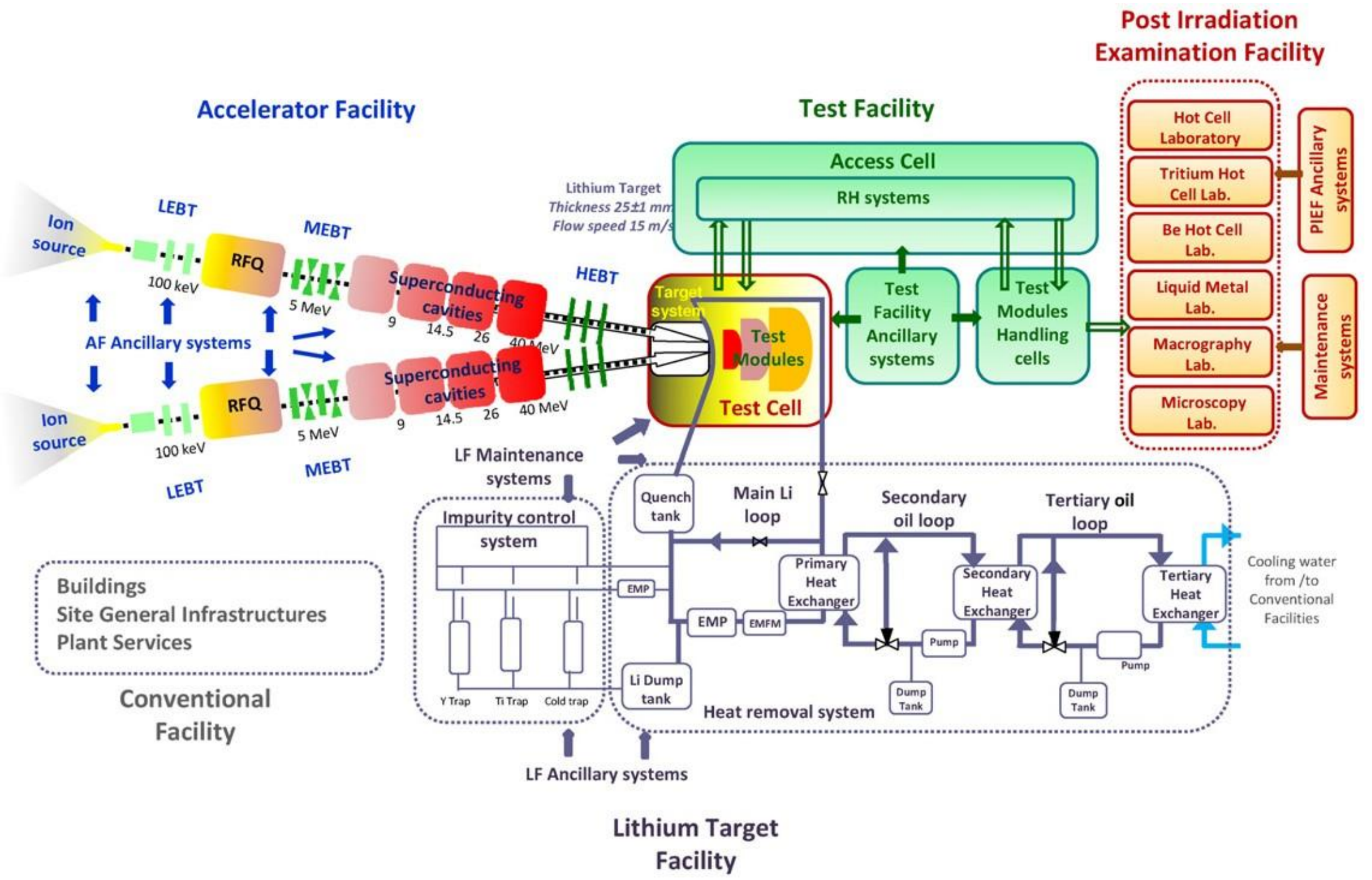

**Figure 2.** Layout of the IFMIF facility.

The accomplishment of the EDA phase in June 2013, exactly within the six years allocated, is intimately linked with the present findings obtained by the validation activities, which, albeit on-going at the time of the release of the report, allowed the definition of the design to be consolidated by the construction and operation of prototypes. A list of all the documents generated is available in figure 3.

The IIEDR is composed of five major elements: (1) the 'executive summary'; (2) the 'IFMIF plant design description' (PDD), that summarises the content of the full IIEDR consisting of more than 100 technical reports; (3) a careful cost and schedule report, based on the experience gained with the construction of prototypes during the EVA phase and the analysis of recognised Japanese and European engineering companies; (4) annexes to the PDD; and (5) 34 detailed design description documents (DDDs) of all the sub-systems supporting the PDD.

Various improvements in the design have been implemented during the EVEDA phase, with the most relevant ones being: (1) the Alvarez-type drift tube linac (DTL) in the AF has been replaced by a superconducting radio-frequency (RF) linac, and consequently the RF system has been simplified accordingly using well-established techniques; (2) the configuration of the test cell (TC) evolved as in the present design, where the irradiation modules no longer have a shielding function and are thus detached from the shielding block, which improves the irradiation flexibility and the reliability of the remote handling (RH) equipment and reduces its costs; (3) the quench tank (QT) of the lithium loop, previously inside the TC, has been re-located outside reducing the tritium production rate and simplifying the maintenance processes; (4) the maintenance strategy together with the management of the irradiated samples has been modified to allow a shorter yearly stop of the irradiation operations [24].

### 4.1. IFMIF site and plant configuration

Various generic site assumptions were put in place for developing the engineering design of IFMIF. These assumptions, addressing required space, topography, geotechnical, hydrological and seismic characteristics, meteorological characteristics, sanitary and industrial sewage, water supply, energy and electrical power, fire protection, radioactive materials protection and radioactive waste, were selected in a conservative manner to ensure their validity for extrapolations towards future site specifics. The IFMIF plan site layout is shown in figure 4.

The main building is a four-storey rectangular building which has a dimension of about 137 m long, 111 m wide, 40.5 m high (27 m high above the ground level). A birds-eye view of the main building is shown in figure 5. The occupational load of the main building is estimated to arise from 66 persons (26 operation staff, 30 daily maintenance staff, and 10 experimentators). The main building contains the AF, LF and TF systems and the plant services of CF.

The TC that houses the target assembly (TA) and the TMs is a blind hot cell (4 m long in the beam direction, 2.8 m wide and 4 m deep) with a unique opening at the top. This opening is closed during irradiation periods by two concrete shielding plugs (SPs) 2.5 m high in total. The inner walls of the TC are covered by a closed steel liner [25].

### 4.2. The accelerator facility

Historical concerns regarding the technological feasibility of operating a deuteron accelerator with the targeted beam

**Executive Summary (this document)**

**IFMIF Plant Design Description Document (PDD)**

**Cost Estimate Report**

**PDD Annexes**

**I. Design Guidelines**

- Annex I.1-CAD Guidelines
- Annex I.2- EDA ALARA Process Guideline
- Annex I.3- Fire Protection Guideline
- Annex I.4- Hazard Evaluation Guideline
- Annex I.5- Human Factor Approach Design Guideline
- Annex I.6- IFMIF Generic Site Assumptions Guideline
- Annex I.7- IFMIF Materials Library
- Annex I.8- Liquid Metal Handling Guideline
- Annex I.9- RAM Guidelines
- Annex I.10- Remote Handling Design Guideline
- Annex I.11- Safety Specifications Guideline
- Annex I.12- Tritium Safe Handling Guideline

**II. PBS and Interface Management System**

**III. 3D MockUp**

**IV. Compilation of Safety-related reports**

- Annex IV.1-Safety Important Class (SIC)
- Annex IV.2- Confinement Strategy
- Annex IV.3- Tritium Source Term
- Annex IV.4- AF Safety Report
- Annex IV.5- LF Safety Report
- Annex IV.6- CF Safety Report
- Annex IV.7- TF FMECA
- Annex IV.8- TF Safety Requirements
- Annex IV.9- PIES

**V. RAMI**

**VI. Licensing Scenarios**

**VII. IFMIF Beam Dynamics**

**SYSTEM DESIGN DESCRIPTION DOCUMENTS (DDDs)**

**Test Facility(TF)**

- Test Cell, Access Cell and Test Module Handling Cell
- TF Auxiliaries
- TF Remote Handling
- Test Modules
  - High Flux Test Module (HFTM) - Vertical
  - High Flux Test Module (HFTM) - Horizontal
  - Start-Up and Monitoring Module (STUMM)
  - Creep Fatigue Test Module (CFTM)
  - Tritium Release Test Module (TRTM)
  - Liquid Breeder Validation Module (LBVM)
  - Low Flux Validation Module (LFVM)

**Lithium Facility (LF)**

- Lithium Facility (Integral Target Assembly)
- Target System - Bayonet Concept

**Accelerator Facility (AF)**

- Injector
- Radio Frequency Quadrupole (RFQ)
- Medium Energy Beam Transfer (MEBT)
- Superconducting RF Linacs
- High Energy Beam Transfer (HEBT)
- Radio Frequency (RF) Power
- Diagnostics
- Beam Dumps
- Accelerator Auxiliaries

**Conventional Facility (CF)**

- Building and Site Infrastructure
- Electrical Power System
- Heat Rejection System
- Heating, Ventilation and Air Conditioning System (HVAC)
- Service Gas System
- Service Water System
- Solid Waste Treatment System
- Liquid Waste Treatment System
- Exhaust Gas Processing System
- Fire Protection System
- Radiation Monitoring System
- Central Control & Common Instrumentation
- Access Control & Security System

**Figure 3.** List of documents of the IIEDR.

performance (running high current beam in CW was the main concern), were rebutted early last decade with the successful commission and operation of the low energy demonstration accelerator (LEDA) in Los Alamos. LEDA, that was the validating prototype of the Accelerator Production of Tritium (APT) project that aimed to accelerate in CW a beam of protons above 1 GeV, successfully produced 100 mA beam current at 6.7 MeV with 99.7% duty cycle for long periods [26].

Each of the two symmetric linacs of IFMIF produces deuteron beams of 125 mA in CW at 40 MeV (see figure 6).

The injector implements the 2.45 GHz and the 875 g electro-cyclotron resonance concept of Chalk River [27] (and successfully operated in SILHI) [28] at 140 mA and 100 kV with a 5 electrode beam extraction system. Two boron nitride disks, typically with low outgassing rates, protect the entrance of the waveguide and the plasma electrode from ion bombardment and help mitigate space charge phenomena. The extracted beam is matched to the radiofrequency quadrupole (RFQ) entrance thanks to a dual solenoid focusing scheme; in turn, the transverse emittance values at the output of the low energy beam transport (LEBT) shall be <0.3 $\pi$ mm mrad [29] and 95% $D^+$ fraction to ensure a transmission >90% at the 5 MeV output of the RFQ.

The RFQ follows the four vanes design [30] successfully operated in Europe in IPHI and TRASCO [31] accelerating the beam to 5 MeV along its 9.8 m length. The shortcomings at low energies due to the space charge effects led to choosing the high input energy of 100 keV with the aforementioned challenging emittance values that will keep losses below 10% until the end of the 'gentle buncher' and below $10^{-6}$ in the high energy part (activation by deuterons, with significantly higher inelastic cross sections than protons, will be within hands-on maintenance limits) [32]. The validation of the tuning and stabilisation procedures were established following low power tests on an aluminium real-scale RFQ [33], which determined the mode spectra and the electric field distribution with the bead pulling technique based on Slater perturbation theory [34].

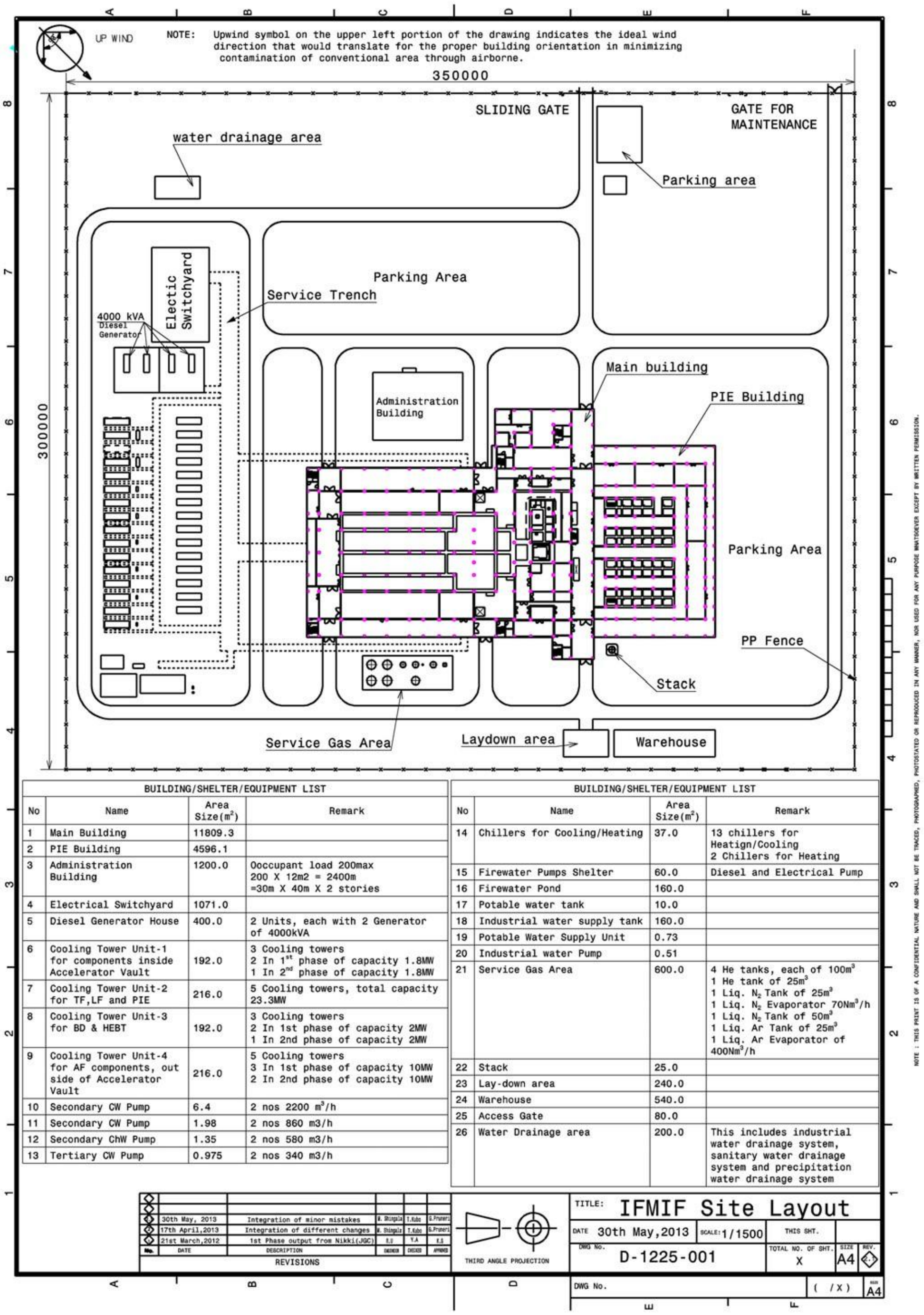

BUILDING/SHELTER/EQUIPMENT LIST

| No | Name | Area Size($m^2$) | Remark |
|---|---|---|---|
| 1 | Main Building | 11809.3 | |
| 2 | PIE Building | 4596.1 | |
| 3 | Administration Building | 1200.0 | Ooccupant load 200max 200 X 12m2 = 2400m =30m X 40m X 2 stories |
| 4 | Electrical Switchyard | 1071.0 | |
| 5 | Diesel Generator House | 400.0 | 2 Units, each with 2 Generator of 4000kVA |
| 6 | Cooling Tower Unit-1 for components inside Accelerator Vault | 192.0 | 3 Cooling towers 2 In 1st phase of capacity 1.8MW 1 In 2nd phase of capacity 1.8MW |
| 7 | Cooling Tower Unit-2 for TF,LF and PIE | 216.0 | 5 Cooling towers, total capacity 23.3MW |
| 8 | Cooling Tower Unit-3 for BD & HEBT | 192.0 | 3 Cooling towers 2 In 1st phase of capacity 2MW 1 In 2nd phase of capacity 2MW |
| 9 | Cooling Tower Unit-4 for AF components, out side of Accelerator Vault | 216.0 | 5 Cooling towers 3 In 1st phase of capacity 10MW 2 In 2nd phase of capacity 10MW |
| 10 | Secondary CW Pump | 6.4 | 2 nos 2200 $m^3$/h |
| 11 | Secondary CW Pump | 1.98 | 2 nos 860 m3/h |
| 12 | Secondary ChW Pump | 1.35 | 2 nos 580 m3/h |
| 13 | Tertiary CW Pump | 0.975 | 2 nos 340 m3/h |
| 14 | Chillers for Cooling/Heating | 37.0 | 13 chillers for Heatign/Cooling 2 Chillers for Heating |
| 15 | Firewater Pumps Shelter | 60.0 | Diesel and Electrical Pump |
| 16 | Firewater Pond | 160.0 | |
| 17 | Potable water tank | 10.0 | |
| 18 | Industrial water supply tank | 160.0 | |
| 19 | Potable Water Supply Unit | 0.73 | |
| 20 | Industrial water Pump | 0.51 | |
| 21 | Service Gas Area | 600.0 | 4 He tanks, each of 100$m^3$ 1 He tank of 25$m^3$ 1 Liq. $N_2$ Tank of 25$m^3$ 1 Liq. $N_2$ Evaporator 70N$m^3$/h 1 Liq. $N_2$ Tank of 50$m^3$ 1 Liq. Ar Tank of 25$m^3$ 1 Liq. Ar Evaporator of 400N$m^3$/h |
| 22 | Stack | 25.0 | |
| 23 | Lay-down area | 240.0 | |
| 24 | Warehouse | 540.0 | |
| 25 | Access Gate | 80.0 | |
| 26 | Water Drainage area | 200.0 | This includes industrial water drainage system, sanitary water drainage system and precipitation water drainage system |

**Figure 4.** Plan site layout.

A medium energy beam transport (MEBT) will realise the transverse and longitudinal beam matching conditions of the RFQ output to the superconducting radio frequency (SRF) input. The compact design minimises the impact of the high space charge while keeping a high flexibility in the beam handling. Five magnets combining quadrupole focusing coils and dipole vertical and horizontal correctors [36], and two 5-gap IH-cavity [37] re-bunchers [38] are assembled in around 2 m. In addition, two scrapers with four movable jaws, each one interleaved between the first three magnets, will stop the beam halo and potential out-of-energy particles coming from the RFQ; this will improve the beam quality and protect sensitive downstream systems like the SRF linac. Each jaw is capable of withstanding a beam power of up to 500 W (2 kW per scraper) [39].

The baseline configuration defined in previous concepts for the deuteron beam acceleration from 5 to 40 MeV relied on a conventional structure, the Alvarez-type DTL. However, considering that all structures of this type have been developed for low intensity projects and operated in a pulse mode at rather low duty cycles (spallation neutron source (SNS) at Oak Ridge, J-PARC at Tokai, or LINAC4 at Cern), the extrapolation to the operation mode of IFMIF, which has to accelerate a high intensity beam in CW mode, was judged as a technological challenge. Thus, during the EVEDA phase, an alternative solution using superconducting half-wave resonators (HWRs) was explored and eventually adopted [24].

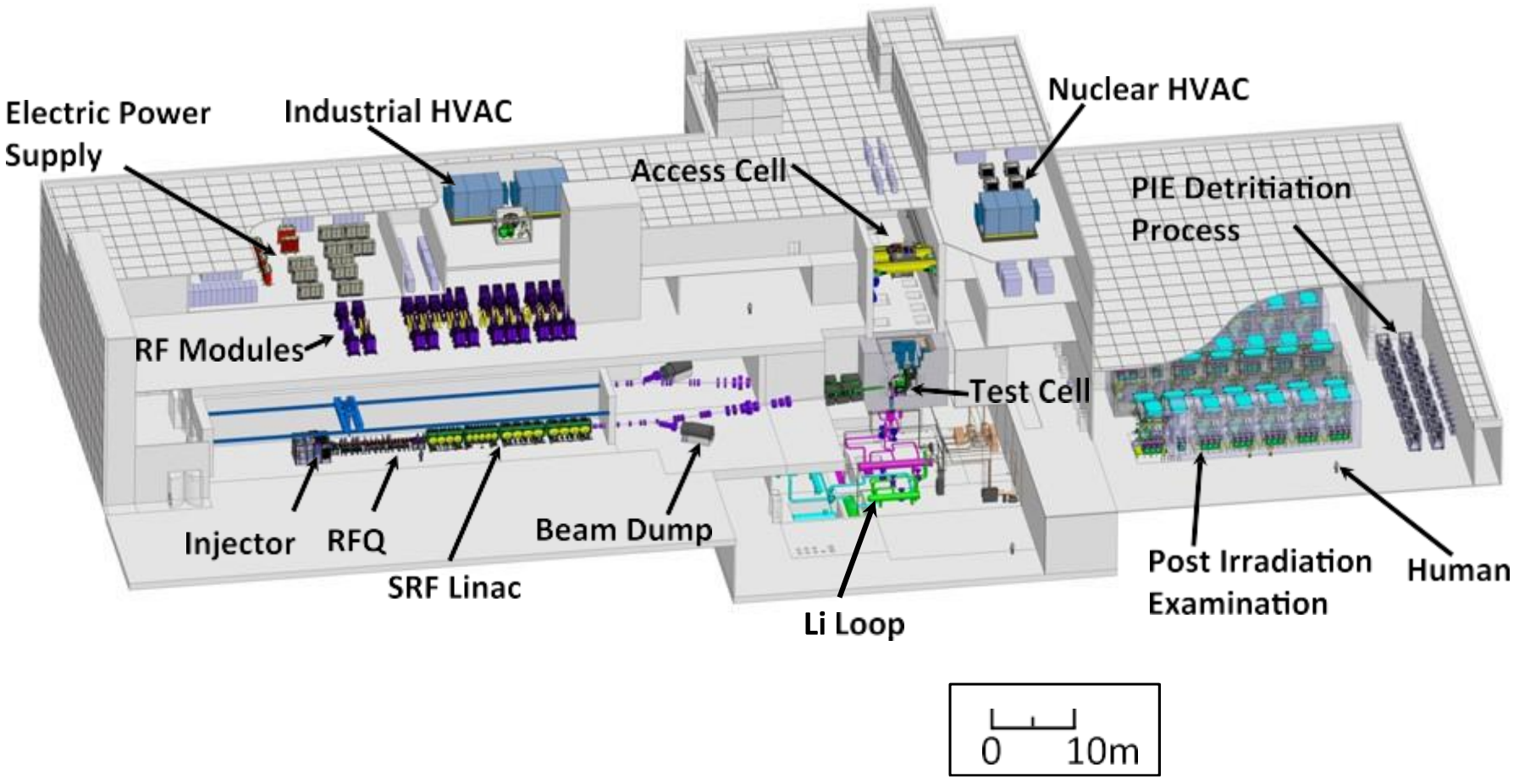

**Figure 5.** Artistic bird's eye view of the IFMIF's main building.

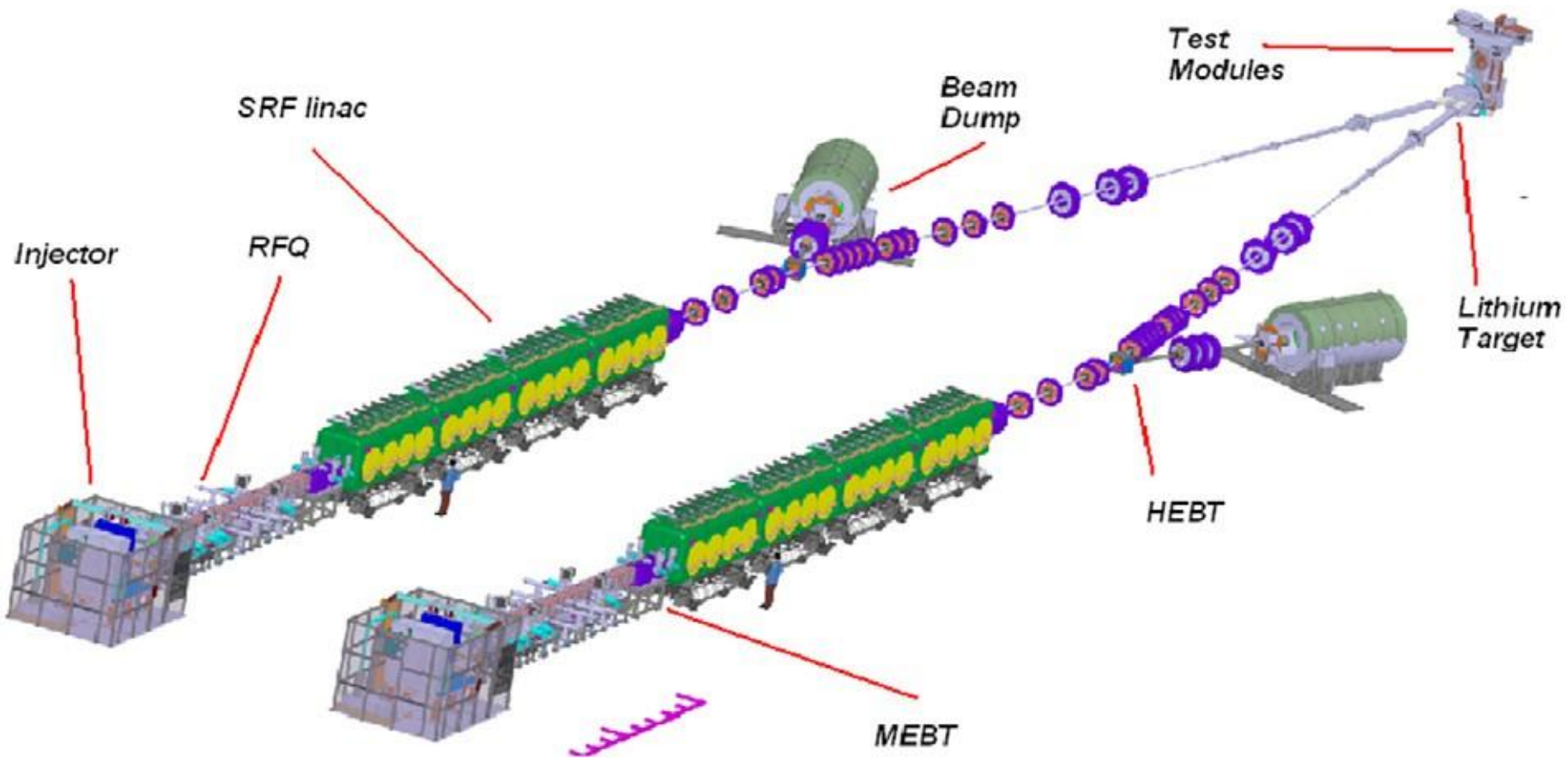

**Figure 6.** 3D Schematic view of IFMIF accelerators, the lithium target and TMs are also indicated.

The chosen configuration in IFMIF is based on a ∼22.7 m long linac, consisting of four consecutive cryomodules. The acceleration of the beam is made by means of RF fields produced in the superconducting HWRs (2-gap cavities at 175 Hz, 4.5 MV $m^{-1}$). The resonant frequency of the cavities is adjusted precisely by using a mechanical tuner (range +30 kHz, resolution 200 Hz). RF couplers provide 200 kW maximum in travelling wave (TW) mode to the HWR. The beam focusing and orbit corrections are performed by sets of superconducting solenoids/steerers and beam position monitors (BPMs) and cryogenic $\mu$-loss monitors, located interleaved with the HWR cavities. The cryostat maintains the superconducting elements below 4.5 K, keeps the internal components under vacuum and insulates them from ambient temperature, pressure and the earth's magnetic field.

The superconducting solution for the accelerator portion of IFMIF offered two main advantages compared with the copper Alvarez-type DTL: (1) linac length reduction (∼10 m) and (2) electrical power saving (∼6 MW) with a positive impact in operational costs. This approach aligned with the technical solution adopted for APT with its 100 MW in CW with 100 mA proton beam [40]. Furthermore, the technological risks linked with the use of CW DTL in the beam nominal conditions were possibly higher than the superconducting choice driven by the growing maturity of the technology since a considerable experience in superconducting low-$\beta$ (the ratio of particle speed to that of light) existed [41] for heavy ions since the 1990s. These operate in CW mode of the independently-phased superconducting cavity linac (ISCL) type, consisting of short (2–4) gaps, low frequency (<200 MHz) accelerating cavities, thus conceptually similar to the proposed IFMIF HWR linac (2-gap cavities at 175 MHz) [42, 59]. In the existing machines, the most widely used resonator type is the quarter-wave resonator (QWR), preferred for its relatively low cost, easy mechanical assembly and high performance at low-$\beta$; however, a significant drawback of this structure is given by the asymmetry of its shape, which can cause undesired beam steering. The HWR approach is similar to QWR but their intrinsic symmetry cancels the QWR steering effect completely. This makes HWR suitable for high current applications with low-$\beta$ beams, keeping most of the QWR virtues without their main drawback. The cavity gradient, the voltage and the phase stability required by IFMIF, as well as the cryostat technology, have been demonstrated [43]. The most recent QWR ISCL, built at TRIUMF, operates normally at peak fields of 35 MV m$^{-1}$, compared with the 4.5 MV m$^{-1}$ of IFMIF HWR ones.

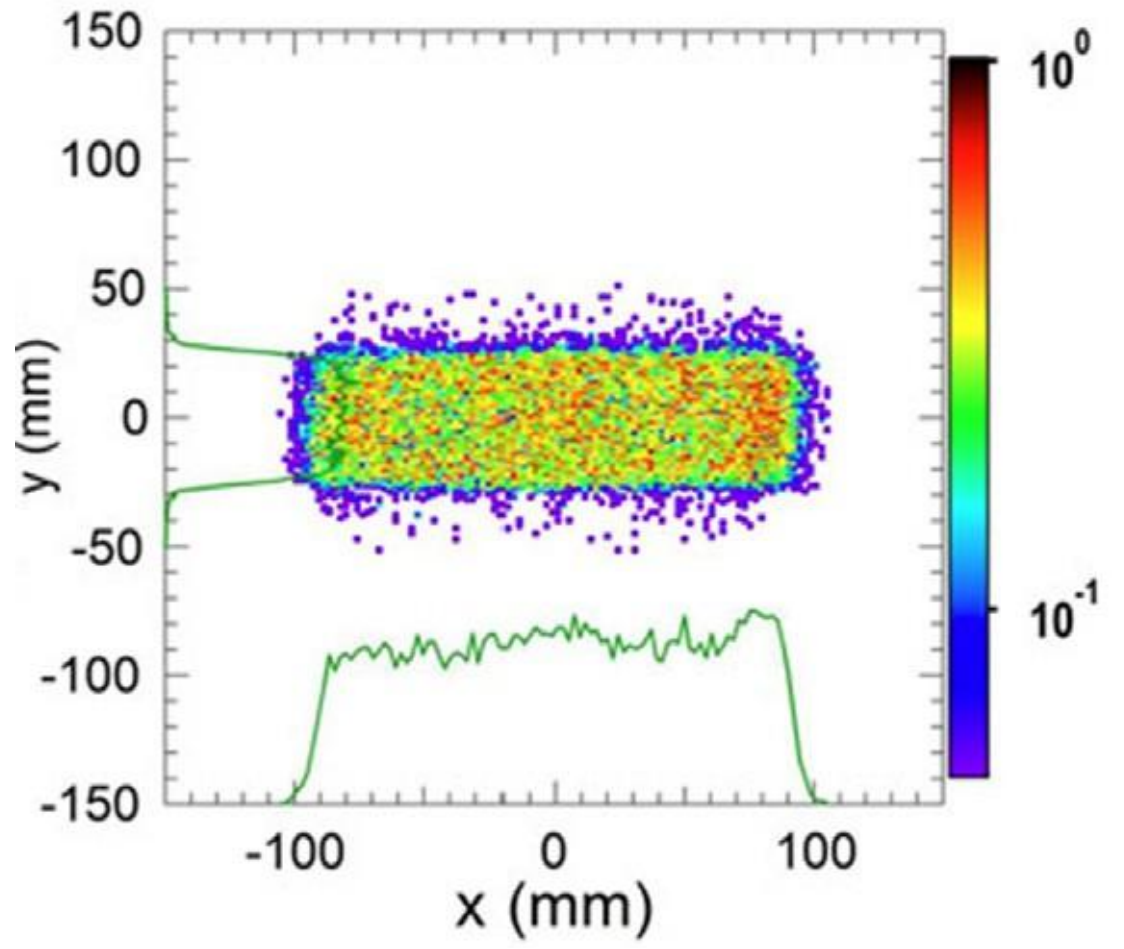

**Figure 7.** Beam density in real space at the lithium target footprint of 200 mm × 50 mm. The density profile projected on both axes is also shown [44].

The objective of the high energy beam transport (HEBT) line is to transport and properly focus the 40 MeV beam coming out from the SRF linac in order to achieve a beam footprint at the liquid lithium target following stringent constraints of dimension and homogeneity with: (1) a rectangular shape of 20 cm (H) × 5 cm (V) on the flat top; (2) beam density across the flat top uniform ($\pm$5%); and (3) beyond $\pm$11 cm in horizontal, beam density lower than 0.5 $\mu$A cm$^{-2}$ (see figure 7). Non-linear multipole optics allows the fulfilment of these specifications.

Under normal operating conditions, IFMIF is fed through the commercial grid by two lines of redundant 66 kV electrical power (100% × 2), with a total power of all connected loads estimated to be approximately 90 MVA. The receiving voltage is stepped down to 6.6 kV via power transformers (66/6.6 kV; 3 × 30 MVA), two of them feeding each accelerator and the other also feeding power to medium voltage switchgears of an emergency power system and an electric distribution system. A dedicated electrical switchyard building hosts the corresponding equipment. In case of grid power failure, emergency generators (diesel engines) will feed electric power to the loads classified as SIC-1 and SIC-2 (SIC standing for safety important components). Emergency generators are provided as a redundant system (50% × 2 + 2; 6.6 kV/4000 kVA), and each generator and relevant equipment is installed in a dedicated building, namely the emergency power building.

Beam halo plays a key role in high-current accelerators becoming the main driver for beam losses, which are to be minimised to remain below the convention for hands-on maintenance of 1 mSv h$^{-1}$ at 30 cm distance of the equipment [45]. In general, one refers to the tails outside the beam core as beam halo; a consensus on its definition is not yet achieved though. As early as 1991, it was already suggested that the most important potential cause of beam loss in the planned high-current linacs would be space-charge-induced emittance growth and halo enhancement [46]. Years of controversy on the underpinning physics were overcome thanks to the installation in LEDA, after the success of its initial APT's concept validation scope, of a 52-quadrupole periodic-focusing beam-transport channel at the 6.7 MeV output energy of the RFQ that allows for the understanding of the beam-halo formation. The experimental results support both models of free mismatch energy conversion into beam thermal energy, predicting a maximum emittance growth and particle-core nonlinear parametric resonance, leading to maximum halo amplitude [47]. The beam core/halo different dynamics have recently led to a novel beam matching method for high power accelerators inspired by IFMIF's conditions [44, 48]. Furthermore, a precise determination of their boundaries has been proposed, allowing for the characterisation of the halo and the core independently [49].

Building a prototype of IFMIF's accelerators to validate the performance was perceived as necessary [50] due to the following reasons: (1) specifications for the IFMIF beam are upgraded in current and energy; (2) RFQ frequency in LEDA was 350 MHz compared with 175 MHz for IFMIF; (3) putting in place an additional accelerating stage to reach 9 MeV (first accelerating stage of IFMIF) would provide an important manufacturing experience; and (4) the use of deuterons would teach lessons regarding nuclear safety considerations.

The linear IFMIF prototype accelerator (LIPAc), under installation in Rokkasho as IFMIF's accelerators validation, follows the design of IFMIF up to its first superconducting acceleration stage with 9 MeV beam energy (see figure 8) [51]. With the deuteron injector and the LEBT being successfully commissioned in Rokkasho and the RF system arriving before summer 2015, the 5 MeV deuteron beam out of the RFQ is scheduled during 2016 and the full LIPAc is to be installed within the timeframe of the Broader Approach agreement. Collective phenomena driven by space-charge forces become the main limitation to achieve high intensity beams. In low $\beta$-regions, the beam radial outward Gauss forces prevail over the inward radial Ampere ones, which mutually cancel in the relativistic domain. Thus, space charge repulsive forces are stronger the lower the beam energy is.

The high beam current and relative low energy challenges the beam diagnostic that has demanded their specific research programs [52], becoming part of the validation activities [19] and their installation cloned from IFMIF. The main features are the very high intensity beams which, together with the low energy and small material penetration, preclude the use of any interceptive diagnostics during nominal continuous beam operation. In addition, the compactness of the accelerator to counteract the space charge forces minimises the space available for diagnostics. Clever solutions have come up to solve this scenario. In the MEBT, for crucial diagnostics, innovative proposals like BPMs inside the interpoles of the magnetic yokes [53] or combined ac and fast current transformers [52] are implemented. Along the SRF, $\mu$-loss and position and phase monitors are placed at low temperatures at each period of the lattice, between the solenoids and the cavities. Beam diagnostics will be a cornerstone for the optimisation of the SRF linac operation. Another essential diagnostics is the measurement of the transverse profile during CW operation; two alternatives are being studied and tested, one based on the residual gas fluorescence [54] and one based in residual gas ionisation [55]. These devices would be used for controlling the rectangular profile and the uniformity of the deposited power at the lithium target. Furthermore, diagnostics to control and improve the beam quality have also been developed and will be implemented in a diagnostics plate to be operational in LIPAc [56]. Last but not least, state-of-the-art devices like a bunch length monitor based on gas ionisation, or a slit withstanding the full beam power (5 MW) in pulsed mode for emittance measurements, that have been designed using novel simulation techniques [57], will be operated in LIPAC.

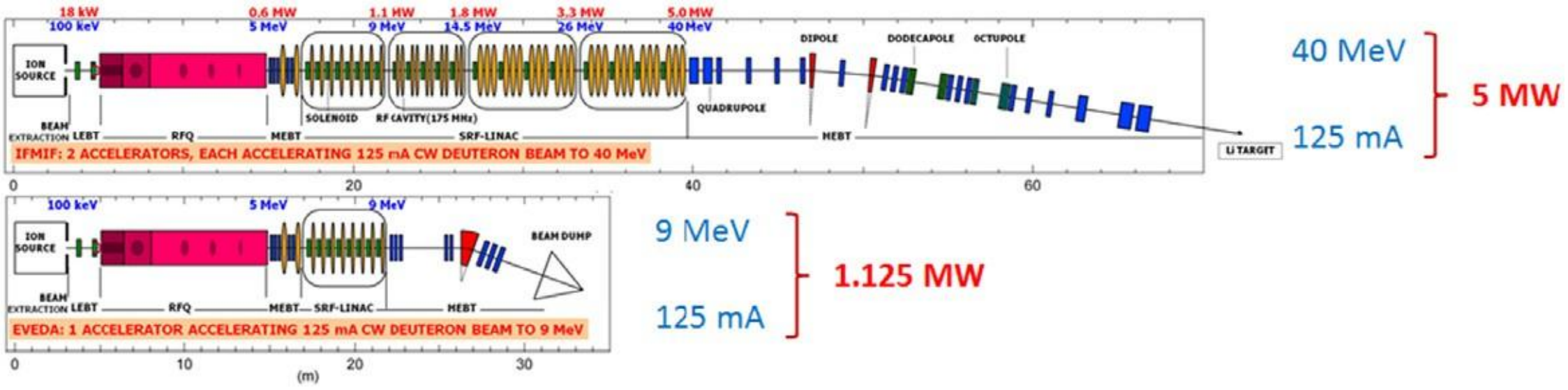

**Figure 8.** Comparison of the layouts of LIPAC and IFMIF's accelerators.

Due to their high beam intensity, LIPAc as well as IFMIF, followed by LEDA, will have the highest beam average power at low energies, and thus for a given beam power, the perveance, the figure of merit of space charge phenomena, will be the highest among the most powerful linacs [58, 61]. The successful operation of LIPAc at 9 MeV will thus validate the concept of IFMIF, as LEDA validated the concept of APT, which had $\times$20 times higher average beam power than IFMIF's [40]. Although LEDA demonstrated the feasibility to operate in CW at 100 mA, and the maturity of today's accelerator technology reliably allows for the construction of facilities with a beam average power in the MW range thanks to the breakthrough achieved at SNS [61], the CW nature of IFMIF's accelerators, and the expected 70% availability of the facility together with its 5 MW beam average power, present an unprecedented challenge that has demanded careful RAMI (reliability, availability, maintainability and inspectability) analysis (see section 6). The RAMI analyses performed for IFMIF's accelerators were based on three main approaches in order to overcome the difficulties of analysing such a novel machine, whose direct comparison with available facilities is not feasible. Firstly, a thorough assessment of the availability of world accelerators was carried out, extrapolating the available experience to what could be expected for IFMIF (taking into account existing differences in machine parameters, technology, environment, maintenance plans, etc). Then, a bottom-up probabilistic analysis allowed us to obtain an estimate of the RAMI performances of the different accelerator systems based on individual components failures. Finally, availability simulations of the whole facility were performed, which included operation, maintenance and beam parameters degradation. The detailed analyses can be found in [60]. These studies allowed for the anticipation of potential problems during commissioning and operation, and technical solutions were found to mitigate them. The results of the analyses concluded that good levels of availability are reachable with adequate redundancies; however, the validation of this RAMI analysis can only happen once the accelerator prototype under construction has operated in CW for a sufficient time.

### 4.3. *The lithium target facility*

The LF, which presents an inventory of about 9 $m^3$ of lithium, provides and conditions the lithium target for the two 40 MeV deuteron beams to generate the required neutron flux from Li(d,xn) nuclear reactions [62, 63]. It is broken down into: (1) the target system, which consists of components situated in the TC and the beam ducts up to the target interface room (TIR); (2) the heat removal system, which consists of the main lithium loop and its dump tank; (3) the impurity control system, which consists of a branch line that extracts a fraction of the lithium from the main loop and re-injects it after purification and impurity analysis; (4) the maintenance system; and (5) ancillary systems, which comprise the control system, the gas supply and exhaust system, the vacuum system, the leak detection and recovery system and the electric power system [64].

The lithium screen serving as the beam target presents two main functions: (1) to react with the deuterons to generate a stable neutron flux in the forward direction; and (2) dissipate the beam power in a continuous manner [65]. A 3D view of the LF is shown in figure 9.

To efficiently fulfil both functions, it shall provide a stable target geometry to the deuteron beam to completely absorb the 10 MW average beam power from both accelerators and protect the thin reduced activation ferritic–martensitic steel (RAFM) backwall plate that channels it. The liquid lithium is shaped and accelerated in the proximity of the beam interaction region by a two-stage reducer nozzle to minimise the transverse velocity components aiming at a laminar flow [65]. In turn, in the beam footprint area, a concave jet of 25 mm thickness with a minimum radius of curvature of 250 mm, builds a centrifugal acceleration of 90 g. This compression raises the boiling point of the flowing lithium guaranteeing a stable liquid phase in Bragg's maximum heat absorption regions (Bragg's peak of deuterons at 40 MeV in lithium is 19 mm). The free surface stability ( $\pm$ 1 mm tolerance is specified) and adequate jet thickness allows us to safely stop the deuteron beam and to limit the fluctuations of the neutron flux in the test specimens (the high flux test (HFT) module [66] is situated at 2 mm nominal distance of the thin backwall plate channelling the lithium). The main parameters of the LF are listed in table 1.

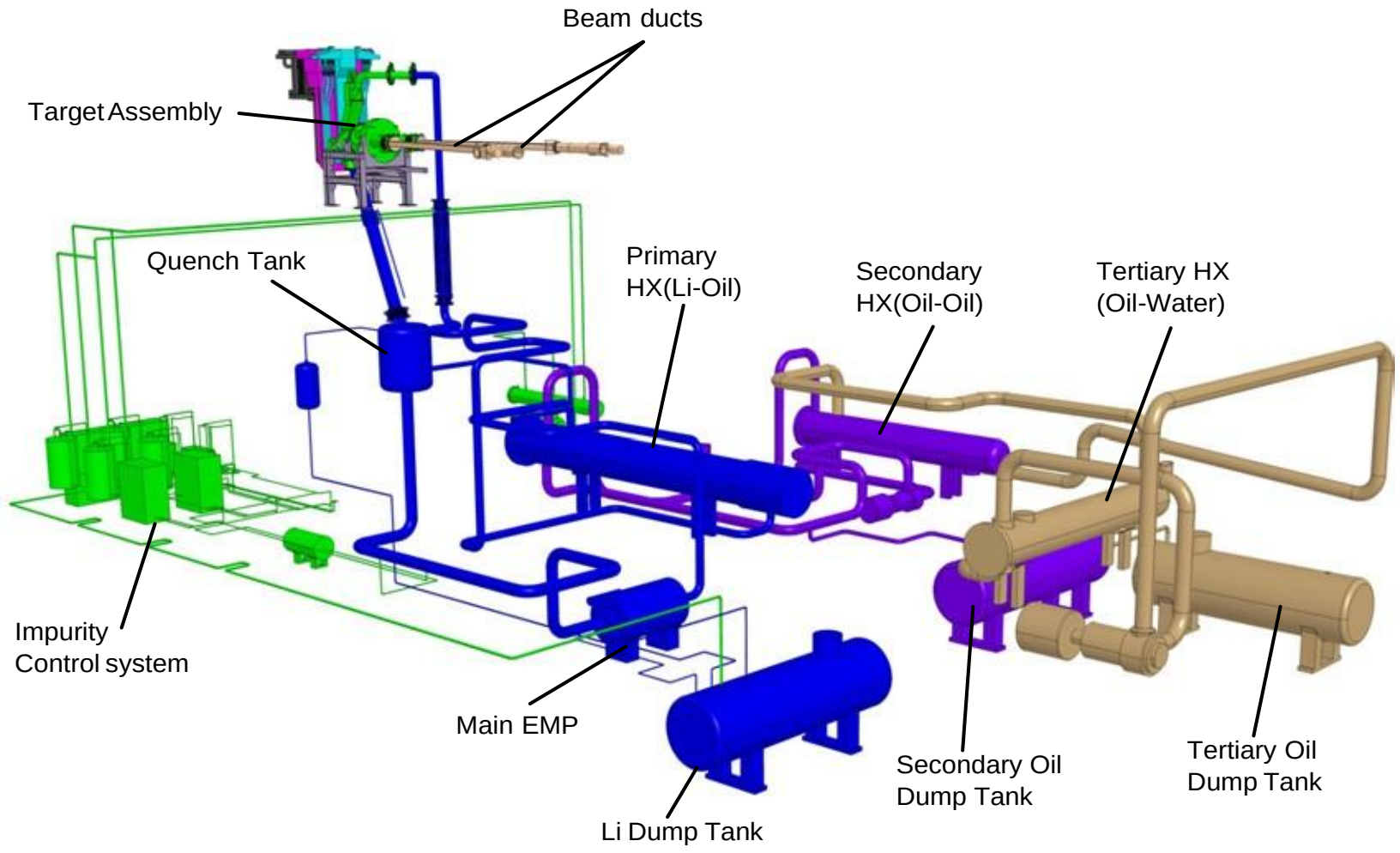

**Figure 9.** Layout of the target facility.

**Table 1.** Major design requirements of the IFMIF LF.

| | |
|---|---|
| **Operation** | |
| Design life | 30 years |
| Facility availability | >94% |
| Scheduled maintenance period | 3 d semi-annual |
| | 20 d annual |
| **Beam on target** | |
| Nominal energy and current | 40 MeV/250 mA |
| Nominal footprint | 200 mm wide $\times$ 50 mm high |
| Injection angle | $\pm 9^\circ$ in horizontal plane |
| Vacuum pressure in ducts | $<10^{-4}$ Pa |
| Vacuum pressure in target chamber | $10^{-3}$ Pa $< p < 10^{-2}$ Pa |
| **Target in footprint** | |
| Li jet thickness | $25 \pm 1$ mm |
| Wave amplitude on surface | <2 mm |
| Distance of Li to HFTM surface | |
| **Lithium chemistry** | |
| Hydrogen isotopes content | <10 wppm |
| Tritium content | <1 wppm |
| O, C, N content | <10 wppm |
| Corrosion rate | <1 $\mu$m per year |
| | <50 $\mu$m per 30 years |

The power density deposited in the flowing lithium is 1 GW m$^{-2}$, a power density which cannot be supported by any solid target. The heat is evacuated with the liquid lithium, which flows at a temperature of 523 K with a nominal speed of 15m s$^{-1}$ exposing its surface to the accelerator high vacuum. The average temperature rise in the liquid is only about 50 K due to (1) the cross flow and its short exposure of 3.3 ms to the two concurrent 5 MW deuteron beams and (2) the high heat capacity of lithium. The beam versus the liquid target interaction has been the subject of careful analysis since FMIT times [67], the results predicting the absence of lithium boiling are backed with recent experiments where proton beam power densities significantly above saturation conditions (>10$^{14}$ W m$^{-2}$) have been reached in flowing lithium at 50m s$^{-1}$ without bubbles nucleation) [68]. Computational fluid dynamics (CFD) calculations have been performed under nominal operational conditions resulting in a maximum temperature $T_{max}$ in the stream of 687.5 K [70]. This temperature is located at the lower edge of the beam footprint, where the lithium has absorbed all the beam energy in a stream wise direction and at Bragg's peak depths, where the centrifugal pressure amounts to about 7 kPa [71, 72], which corresponds to a $T_s$ of 1304 K. Turbulence is expected to enhance the transfer of heat, but most of the turbulent region is away from the region of higher temperatures thus limiting the effect of turbulent diffusion to a slight increase in the wall temperatures near the exit. The maximum temperature in the free surface is 574 K [71], which corresponds to a $P_s$ of $1.3 \times 10^{-4}$ Pa. Thus, at operational pressures of $10^{-3}$ Pa compatible with beam vacuum requirements, the operational margin in the free surface would be of 41 K (see figure 10), which in addition would be increased based on the observations of an increase of pressure in the liquid–gas interphase in experiments with electron beams colliding in a flowing lithium screen [69, 77].

Furthermore, the liquid lithium speed (15 m s$^{-1}$) is too high to allow for the appearance of constructive interferences of pressure waves (maximum possible speeds of 0.5 m s$^{-1}$ from thermal impact or momentum transfer), and at the same time pressure wave amplitudes are damped down by centrifugal pressures (32 Pa maximum pressure driven by beam momentum transfer compared with the centrifugal pressures induced by the concave backwall plate in the order of kPa in Bragg's peak regions) [65]. The free lithium surface exposed to the beams requires an optimal control of the vacuum pressure to maintain the suitable high vacuum conditions in the accelerators and avoid degradation of accelerator components by uncontrolled lithium outgassing. The typical low conductance of beam pipes allows for differential pumping on the beam versus the liquid lithium interaction region resulting in pressures above $10^{-3}$ Pa in the TA compatible with the ultra-high vacuum values demanded in other regions of the accelerator.

The heat removal system is designed to remove the heat deposited by the beams in the target and maintain a defined lithium temperature and flow rate at the TA inlet. It has the flexibility to operate also at intermediate power levels, i.e.

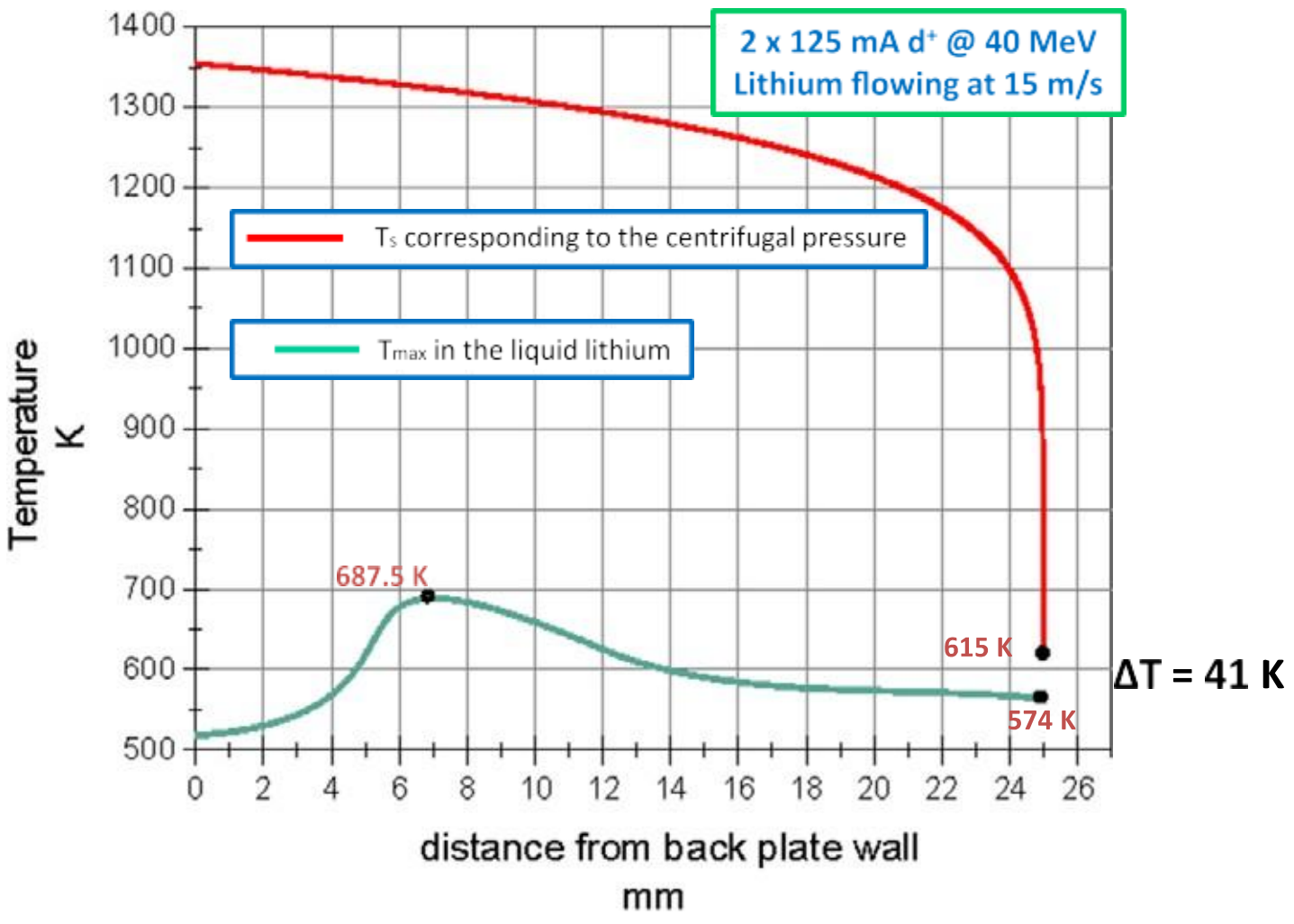

**Figure 10.** $T_{\max}$ envelope in the beam footprint under nominal conditions at different depths (in green) versus $T_s$ corresponding to the centrifugal pressure in the flowing lithium (in red); 615 K corresponds to the beam line nominal pressure of 0.001 Pa [65].

when only one accelerator is in operation or operates at a lower current than nominal and it must also be capable of managing transients during beam start-up or shutdown and trips of one or two accelerators. The nominal inlet temperature at the TA is set to 250 °C. The heat deposited by the beams raises the flowing lithium temperature to 298 °C. The heat removal system of the main lithium loop circulates the 97.5 l $s^{-1}$ lithium flow from the exit of the beam target to a 1.2 $m^3$ QT, where it is slowed down and thermally homogenised before it flows to the electromagnetic pump. The lithium is then cooled back to 523 K by a serial of heat exchangers (HXs) (figure 11 shows the flow diagram for the heat evacuation).

The heat is extracted via two intermediate oil loops and transferred to the plant cooling water. Whereas only one intermediate loop was considered in the IFMIF Comprehensive Design Report (CDR) [18], two intermediate loops have been added to reduce the thermal gradients in the HXs and avoid boiling the water in case of loss of flow. Due to the low thermal capacity of the oil and the moderate thermal gradients, flow rates in the oil loops are rather high requiring large pipe dimensions. Table 2 lists the cooling fluid, flow rates and the temperature ranges of the loops under nominal conditions.

The large volume of the cooling fluids and the high heat capacity of the lithium afford a high thermal stability of the system. Temperature variations due to thermal transients such as beam current fluctuations or trips are slow and the response time for the temperature control is well beyond normal control time constants. The inertia of the systems allows setting a dead time of 100 s before active control is activated by acting on the oil flow through the HX.

There are two drain lines between the main lithium loop and the lithium dump tank (DT) as redundant measures to overcome possible malfunctions in the opening of the gate valves. The temperature of the lithium in the DT is monitored and controlled. The lithium loop operates under vacuum conditions to comply with the accelerator needs, but is filled with argon gas during maintenance.

The QT, which controls the lithium level in the hot leg of the loop, is positioned at a height which avoids cavitation in the inlet of the pump. The QT of the lithium loop, previously included inside the TC, has been re-located outside. Thanks to this change the tritium production is reduced and, in addition, the operations required to exchange the QT, in case of failure, have been simplified.

The impurity control system in the lithium will be done through tailored design cold and hot trap systems; purities of lithium during operation better than 99.9% are expected. The presence of impurities in the flowing lithium not only have implications on nuclear safety, given the radioactive by-products of the Li(d,xn) nuclear reactions [62, 64], but might also have implications on the free surface stability (the presence of gases as well of solid elements in suspension might favour the nucleate boiling). On-line monitoring of impurities will detect trips of impurities over 50 wppm. The impurity control system consists of a branch line, which extracts a fraction of the lithium from the main loop and re-injects it after purification and impurity analysis. The system is designed to condition the lithium after maintenance prior to start-up and control and maintain a defined level of purity. The purification branch contains: (1) cold traps to collect impurities with temperature sensitive solubility, such as oxygen, carbon, beryllium and other corrosion products within 10 wppm; (2) hot traps to specifically capture nitrogen chemically within 10 wppm, which has a very high solubility in lithium and cannot be removed to the required level by cold trapping; and (3) hot traps to extract tritium within 1 wppm by specifically binding all hydrogen isotopes chemically.

The cold trap extracts impurities as binary or ternary compounds based on their solubility. The getter material used is a stainless steel wire mesh at 200 °C. The inflowing lithium with a temperature of 250–300 °C passes through an economiser and is cooled in the trap by an argon cooling circuit. The traps are sized based on the basis of the oxygen source term assuming an initial content of 1360 g corresponding to a concentration of 272 wppm and an annual recontamination of 1280 g due to maintenance operations. Two cold traps are conceived for redundancy, but only one economiser is employed to avoid interruption of the operation in case of failure or insufficient performance. The capacity of the two traps is dimensioned to cover the 30 years of the IFMIF planned lifetime. Given that the initial impurity will be removed during the commissioning phases of the TF, in the absence of activated products, this will be done in a dedicated trap or packing.

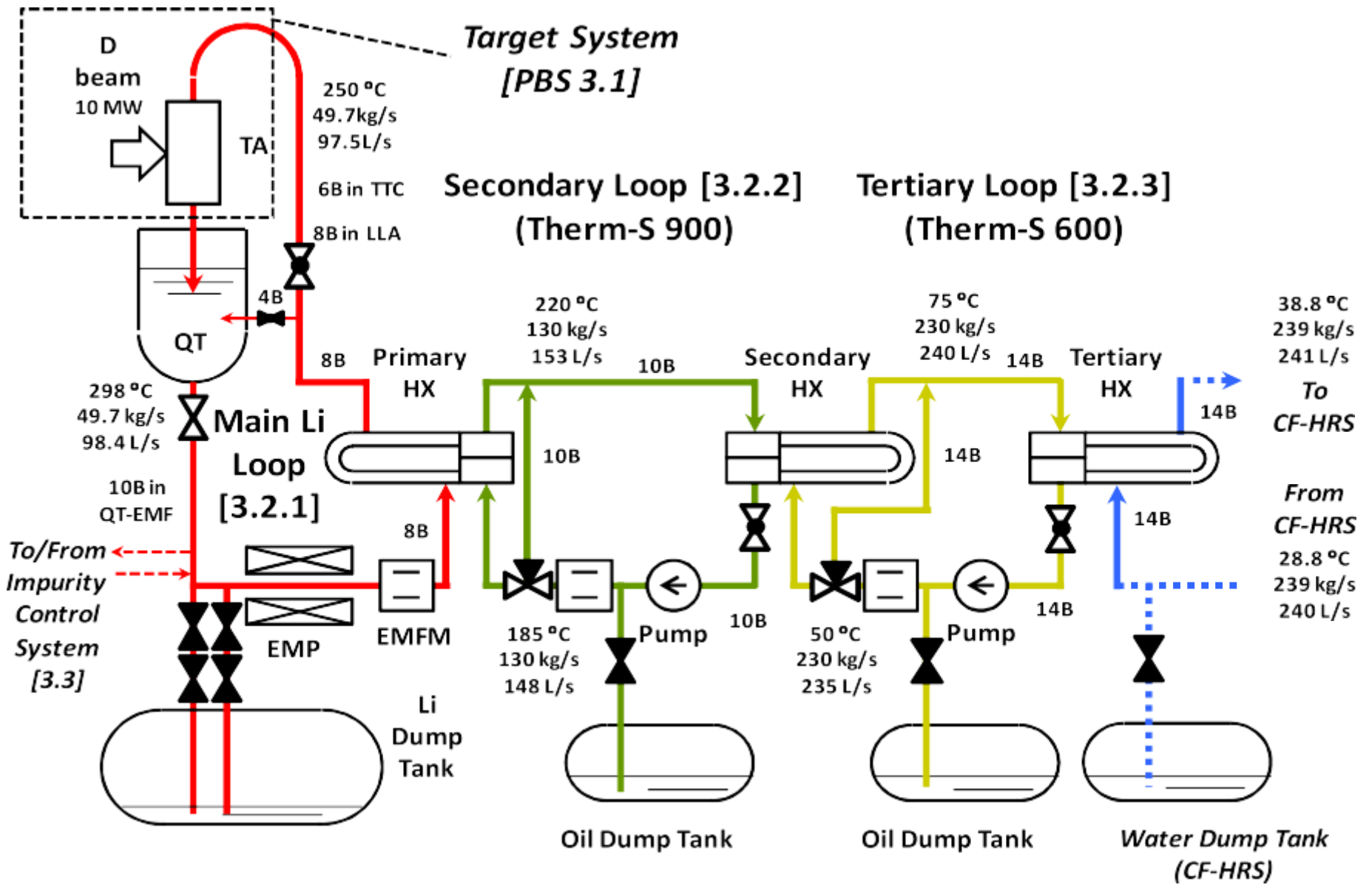

**Figure 11.** Block flow diagram of the heat removal system.

**Table 2.** Summary of heat exchange in the heat removal system (flow velocity $= 15\,\mathrm{m\,s^{-1}}$).

| Section | | | Flow rate | Temperature (°C) | |
|---|---|---|---|---|---|
| HX | Side | Fluid | ($\mathrm{kg\,s^{-1}}$) | Inlet | Outlet |
| Primary HX | Shell | Lithium | 49.7 | 298 | 250 |
| | Tube | Organic oil (ThermS-900) | 130 | 185 | 220 |
| Secondary HX | Tube | | | 220 | 185 |
| | Shell | Organic oil (ThermS-600) | 230 | 50 | 75 |
| Tertiary HX | Tube | | | 75 | 50 |
| | Shell | Water | 239 | 28.8 | 38.8 |

A hot trap extracts nitrogen, which has a very high solubility in lithium and cannot be reduced to the required concentration level by cold trapping. Titanium is used as the getter material to form TiN [74, 75]. In order to provide sufficient reactivity, the trap is operated at 600 °C nominal temperature. The inflowing lithium is preheated in a special heater and flows through an economiser to recover the heat in the outflow. The trap is sized based on the nitrogen source term assuming an initial content of 1500 g and an annual recontamination of 140 g, resulting in an annual increase of nitrogen concentration of about 30 wppm. The capacity of the two traps is dimensioned to cover the IFMIF lifetime. These traps, though technologically feasible, have not been designed in detail in this accomplished EDA phase. Initial purification during the commissioning is possibly recommended in a dedicated trap. The use of Fe–Ti alloys in two stages, by re-injecting the purified lithium <100 wppm nitrogen in a new purification loop, overcomes the saturation observed at levels above the target values.

A hot trap extracts hydrogen isotopes within 10 wppm, which are initially contained inside lithium and which are continuously produced in the target during beam operation. The getter material is yttrium, which shows a higher affinity to hydrogen than lithium [85]. The trap operates in the temperature range of 280–300 °C and thus requires only moderate temperature control afforded by the trace heating of the inlet piping and the trap. The initial content of hydrogen is assumed to be 720 g in the design basis, which is to be removed during the commissioning phase. In this phase, anyway, the hydrogen hot trap is to be employed only after the nitrogen hot trap has fulfilled its duty, in other words, when lithium has been already purified from nitrogen. Yttrium getter, in fact, is able to adsorb nitrogen too, forming a compound (YN) which is more stable than yttrium hydride and could therefore lead to its saturation and consequent loss of efficiency [81]. The hydrogen hot traps are sized to extract an annual production of hydrogen isotopes (83.3 mol, whose 93% is deuterium and 4% is tritium) and are scheduled to be replaced annually during the maintenance period. The traps are considered redundant to assure continuation of the operation in case of failure. Operation of the hydrogen trap during beam operation is essential to limit the tritium level in the target. Given the high initial amount of non-active hydrogen, it is recommended to getter it in a dedicated trap or packing during the commissioning phase.

The impurity control system holds a monitoring branch, which in turn contains lithium samplers to collect and freeze lithium samples for off-line analysis; the unit is arranged to allow sampling and extraction of the sampler during beam operation and stations monitoring the impurity content online: (1) a plugging meter to determine with high precision the lithium freezing point, which is indicative of the overall impurity content and the oxygen content in particular; (2) a resistivity meter to measure the electric resistance of the lithium, which is indicative of the overall impurity content and of the nitrogen content; and (3) a hydrogen sensor which measures the hydrogen isotope content based on permeation through a thin membrane and is able to discriminate between the different isotopes.

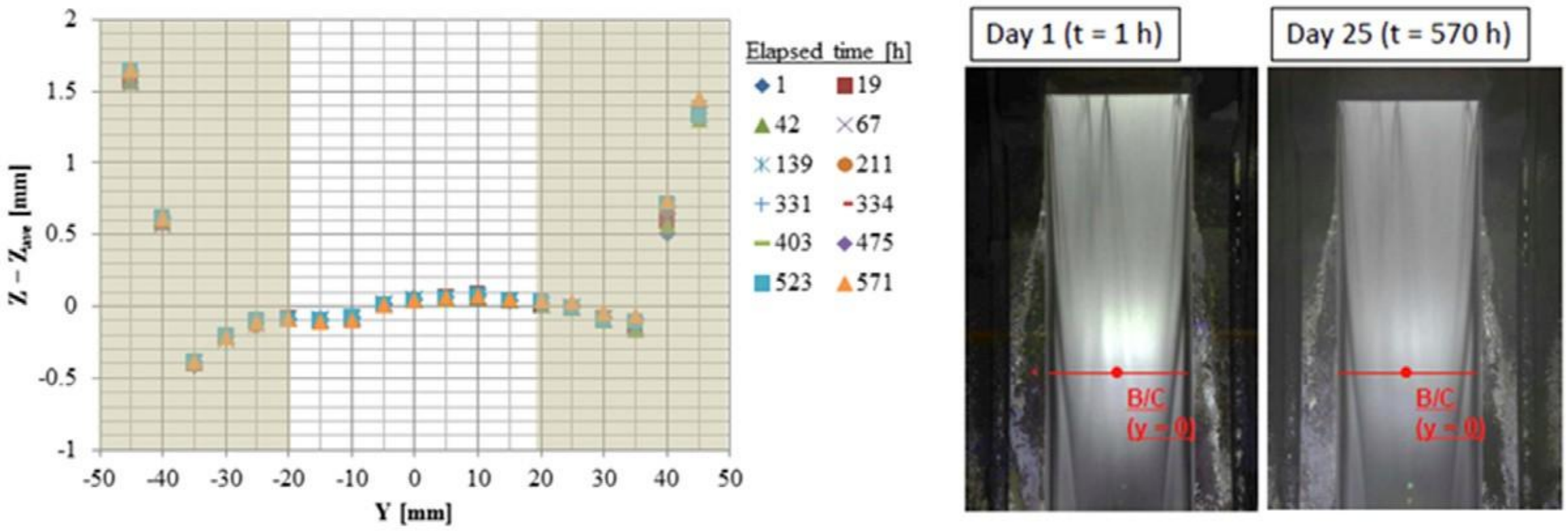

**Figure 12.** Evolution of amplitudes in the free surface of ELTL flowing continuously 25 d under nominal conditions (15 m $s^{-1}$ and 523 K).

A corrosion limit of about 1 $\mu$m per year has been set for the backwall plate and 50 $\mu$m in 30 years in loop conduits. This requirement is assumed to be achieved by limiting the flow velocity and lithium temperature and by controlling and maintaining the chemistry within defined tolerances to limit corrosion/erosion of the structural materials and the dissolved nuclear inventory. The erosion rate follows a parabolic law with fluid velocity [76, 78] and is enhanced by high nitrogen content by forming $Li_9CrN_5$ ternary compound [78]. A target has been set to limit the concentration of nitrogen and oxygen (which plays a role in the formation of the ternary compounds) in the lithium to <10 wppm each. This value is perceived as safely conservative since the measured corrosion rates of metals exposed to lithium drop below 723 K [79], and 80 wppm seems to be the threshold for the formation of $LiN_3$ [80], an indispensable step in its chemistry. In turn, deuteron and neutron interaction with lithium generates radioactive products, essentially tritium and $^{7}$Be impurities and dissolved corrosion products become activated when transported through the beam footprint. Limiting the nuclear inventory in the lithium in view of accident mitigation and managing the distribution of gamma-emitters is required to assure accessibility to the loop area during maintenance operations. A target to limit the tritium concentration in the lithium to <1 wppm has therefore been set.

Under the validation activities, the EVEDA lithium test loop (ELTL), physically equivalent to the loop in the IFMIF plant, was constructed [82]. The main validation targets were successfully accomplished, namely, free surface amplitudes within target tolerances of ±1 mm at nominal operational conditions, free surface interferometry diagnostics that can allow on-line monitoring during beam operation [83] and long term operation stability (>1000 h) under nominal flow conditions (see figure 12) [84]. Whereas cold traps have been successfully used in the ELTL, hot traps for nitrogen and hydrogen gettering were not installed.

In turn, corrosion/erosion phenomena is under study in LIFUS6, a lithium facility designed and constructed [73] to understand the corrosion/erosion phenomena induced by the presence of nitrogen solved in flowing lithium under the operational conditions of IFMIF (15 m $s^{-1}$ and 523 K).

An alternative TA, which allows for the reduction of waste and shortening the maintenance time for the LF components in the TC [86] is also considered, although it presents a higher complexity. The back wall is conceived as a removable part (backwall plate) connected to the TA body by a flange sealed with a rectangular metal gasket and fixed by a movable bayonet system. The gasket is retained in the grooved backwall plate by a number of clips to secure it in place during mounting and dismounting. The backwall plate slides vertically into its position using a special jig. It can be mounted and dismounted by remote operation allowing the TA to stay in place during the backwall plate replacement and includes a gasket groove equipped with a leak detection wire. A 3D view of its configuration is shown in figure 13.

### *4.4. The test facility*

The TF [87] includes the systems required to accommodate the TMs under controlled environment and conditions for irradiation, as well as all the systems required for their assembly and disassembly and sending the irradiated specimens to the PIEF. The TF comprises all equipment, primary heat removal systems, purge gas systems and handling facilities for an accurate and safe positioning and handling of the specimen, modules and target during beam operation and maintenance. It is formed by the TMs, TC, access cell (AC), test module handling cells (TMHCs), test facility ancillary systems (TFAS), and RH systems (the TFAS are also known as test facility utility rooms in the reports of previous IFMIF phases). Figure 14 shows the 3D view of the TF design in the main building.

The main missions of the TF are: (1) housing the Li(d,xn) reactions; (2) disassembling and assembling of the TM including insertion and extraction of specimens; (3) replacement of TA and TMs; and (4) transportation of the specimens between the TF and PIEF. The maintenance system is shared between the TF and the LF. The TF will provide standard RH systems to remove and to insert the LF components (mainly the TA and lithium pipes) while the design of the LF includes the specific tools for those RH procedures like the bayonet backwall plate. The two key spaces of the TF are: (1) the TC housing the TMs; and (2) the set of hot cells allowing the replacement of the TA and the TMs, the preparation of new modules and the extraction of irradiated specimens.

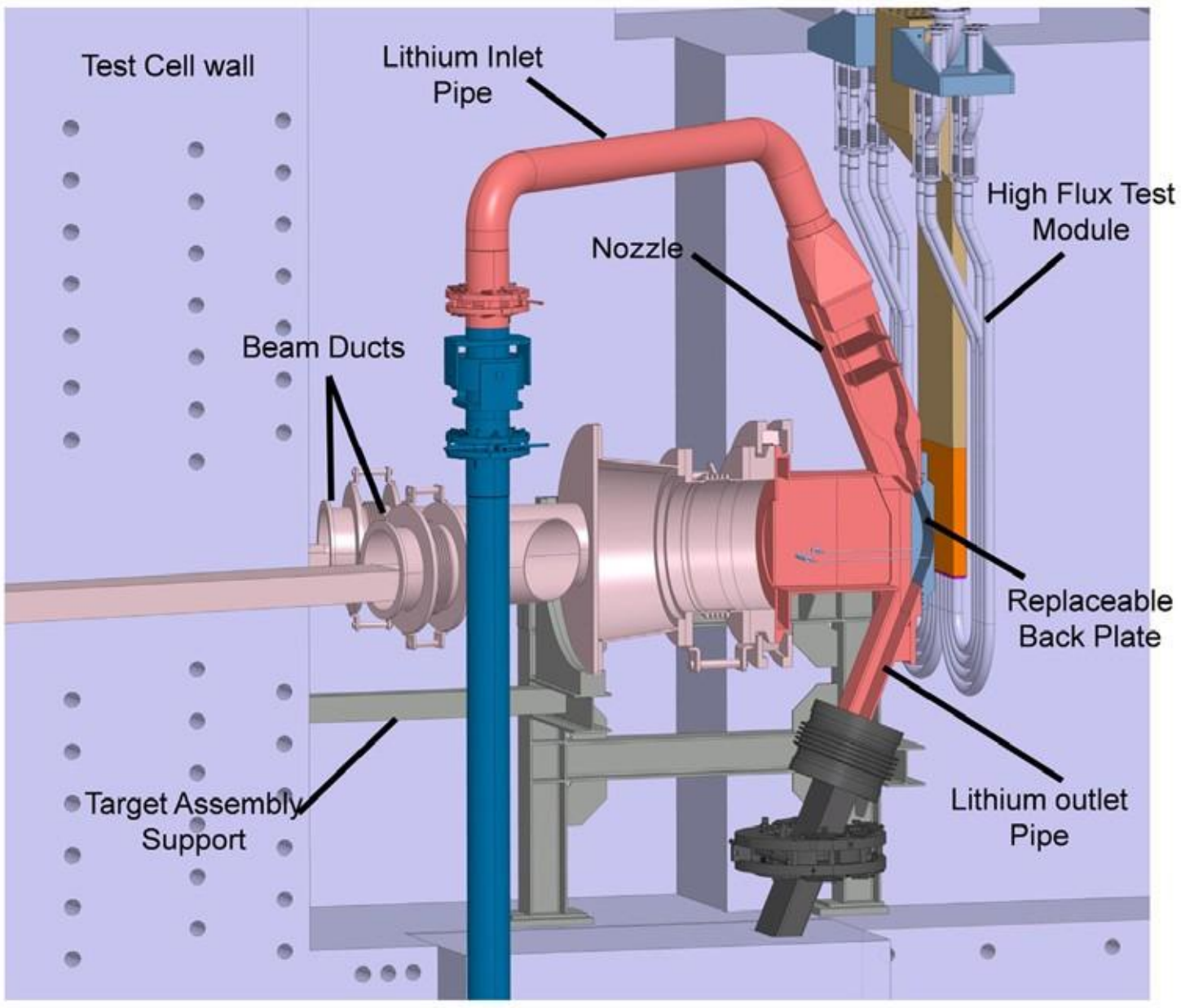

**Figure 13.** Deuteron beams and the lithium target interface region with a bayonet type backwall plate.

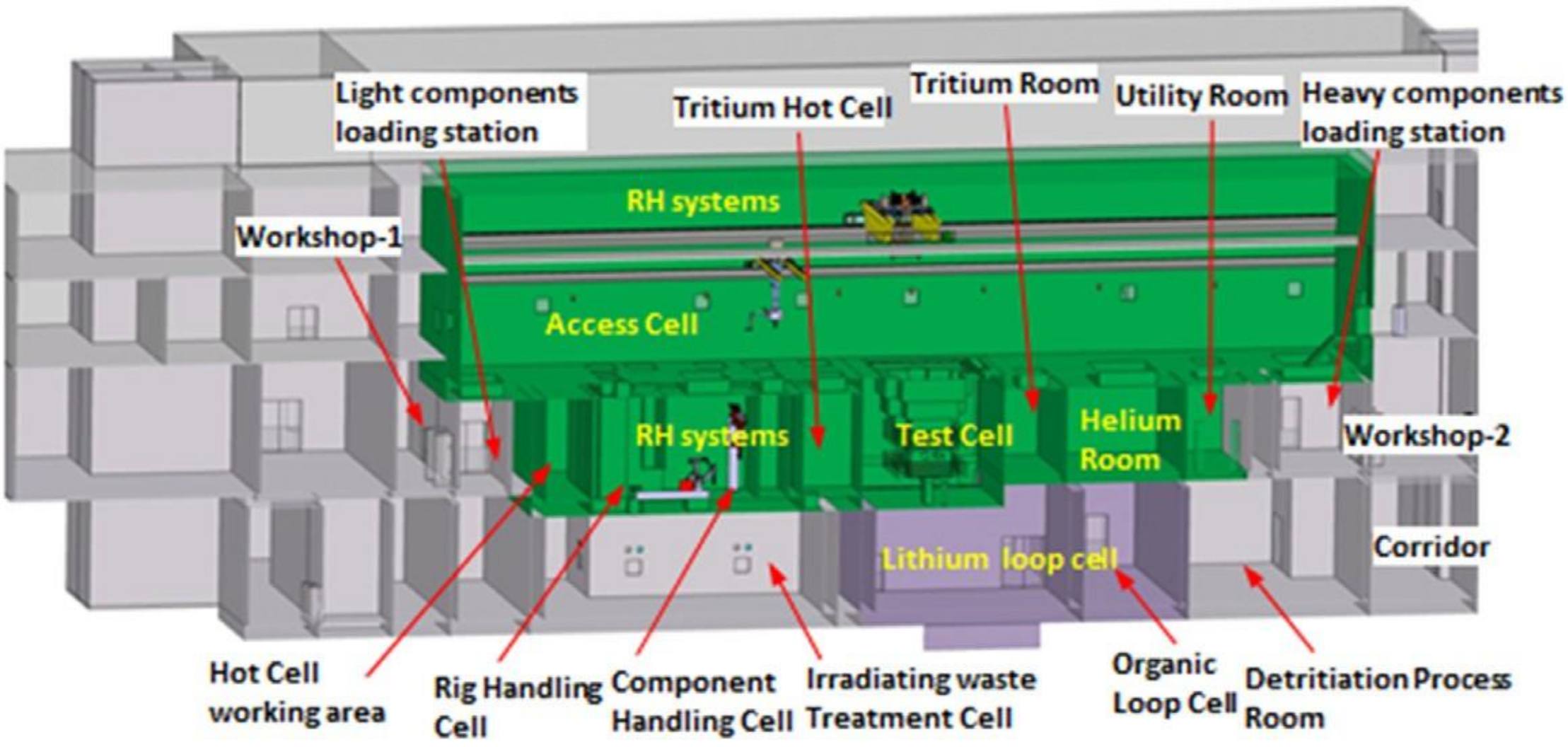

**Figure 14.** Arrangement of the TF in the building. Some rooms that are part of the LF and CF can also be appreciated.

*4.4.1. The test cell.* The TC is a blind hot cell with an opening at the top [87] (see figure 15). The surrounding shielding walls are riveted with a liner which provides, together with the TC upper cover plate, a vacuum tight enclosure to guarantee that an inert atmosphere is maintained during beam operation with a negative $\Delta P$. The liner and biological shielding (BS) are made from concrete and cooled with chilled water. The TC structure serves as a checkpoint for the orientation or fixation of the TC internals in relation to the beam axis. The BS of the TC is completed by the SPs. The top closure of the TC is split into two top SPs. The connections between the TC and the external world for transferring liquids, gases or signals are made through the piping and cable plugs (PCPs).

The two top SPs have to be removed every time access to the TC is granted. The lower shielding plug (LSP) is actively cooled by helium whereas the upper (USP) one is uncooled. Thanks to this split, the maximum weight load for the crane is optimised. In addition, the plug shape corresponding to the TC opening ensures that in case of a load drop no plug would fall into the TC cavity. The BS, in particular the USP and LSP, ensures that even at full beam power the radiation levels in the AC allows for the accessibility of workers.

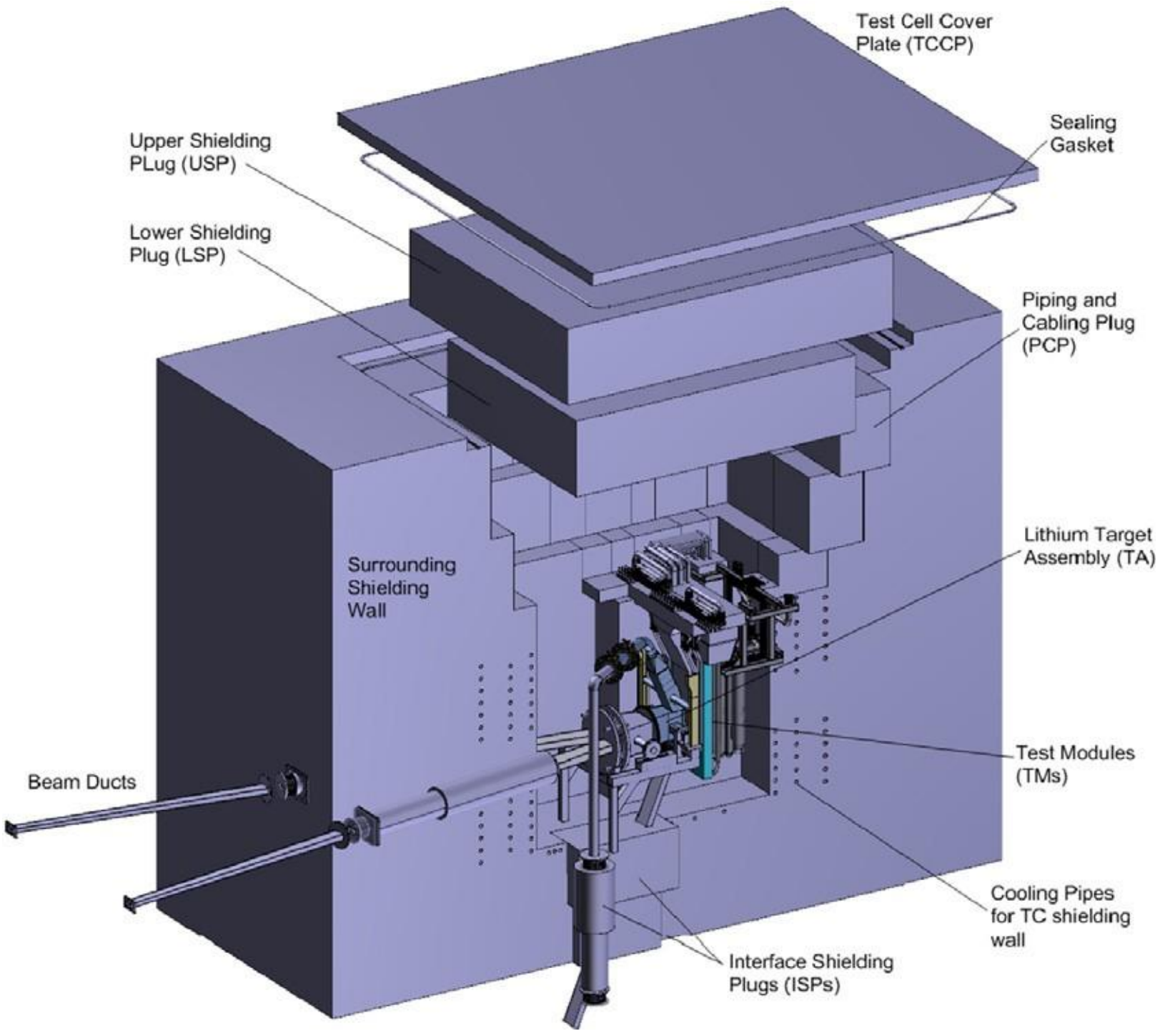

**Figure 15.** TC with internals and penetrations for beam tubes as well as a lithium loop inlet and outlet pipes.

The PCPs remain in position for standard module or target replacement and their shape is designed to minimise radiation streaming. The six PCPs are made of heavy concrete and covered with a stainless steel envelope (see figure 16). The PCPs also bear the functions of: (1) shielding the neutrons and gammas from reaching the AC from laterals; and (2) tightening the TC by being welded at the floor of the AC.

The TMs are supported from the TCs walls, which are parts of the BS allowing their independent operation. The final tightening of the TC is achieved by the TC covering plate. It closes the TC over the SPs, as shown in figure 15. The cover sheet and in particular the sealing against the liner is outside the high dose radiation field.

The liner and cover are designed for an inner sub-pressure of 1 mbar. The free volume of the TC cavity and the entire volume of all gas/helium loops connected to the TC are related in so far as no over pressure of the TC may occur in case of a leak. Therewith, an over pressure design is excluded.

The lithium QT is located directly below the floor of the TC, right under the ceiling of the LF room. The QT is connected to the lithium TA through the lithium outlet pipe, which penetrates the floor of the TC and is surrounded by an interface shielding plug (ISP). In turn, the lithium inlet pipe also penetrates the TC floor through another ISP. The ISPs are designed to minimise the neutron streaming from inside the TC to adjacent rooms. Each of them can be extracted from inside the TC in case of the failure of any of the lithium pipes. While parts of the ISPs are exposed to intense neutrons and gammas inside the TC, active cooling of the ISPs is required. The penetration of the beam ducts through the TC shielding walls also present ISPs; these are installed in the wall between TC and the TIR (see figure 17). The thickness of the wall between the TC and the TIR is designed to be 3.5 m. The concrete walls and floor of the TC are actively cooled with water to remove the nuclear heating due to neutron and gamma during irradiation. It has been estimated conservatively that 5% of the beam power will be released by the shielding concrete, thus 500 kW. These concrete walls are completely covered with an 8 mm thick stainless steel liner, which is kept at 20 °C during irradiation thanks to active cooling with helium. At the insider surface of the TC, the ISPs are welded with the TC liner to tighten the TC. The liner prevents fire accidents through reactions of accidental lithium spills with concrete.

*4.4.2. The test modules.* Three different irradiation areas are foreseen behind the backwall plate in the TC for TMs installation: the HFT region, the medium flux test (MFT) region, and the low flux test (LFT) region [89] (see figure 18). The HFT area provides a total volume of 0.5 l to house specimens at a damage rate of 20–50 dpa fpy$^{-1}$. The MFT area provides a total volume of 6.0 l for the specimens with a damage rate of 1–20 dpa fpy$^{-1}$, and in the LFT area, the yearly accumulated fluencies will range between 0.01 and 1 dpa with a higher available volume. During the irradiation period, all of the components inside the TC are exposed to intense neutron and gamma irradiation. The nuclear heating applied on these components has to be removed continuously, thus all of the TMs will be actively cooled with their own cooling system.

Two different module concepts have been defined in the HFT region: (1) the HFT module vertical layout (HFTM-V) and (2) the HFT module horizontal layout (HFTM-H), which are expected to be arranged inside the TC in different irradiation campaigns. Three different modules have been designed for the MFT area: (1) the creep fatigue test module (CFTM); (2) the tritium release test module (TRTM); and (3) the liquid breeder validation module (LBVM). In addition, neutron spectral shifters (NSS) could be installed. It is to be noted that these three modules cannot be simultaneously located in this area and different configurations will be used in the different irradiation campaigns. The LFT area is capable of housing several containers in which different experiments can be performed and accommodated in the low flux test module (LFTM). Last but not least, a start-up monitoring module (STUMM), only used during the commissioning phase of IFMIF, is also included in the TF [88].

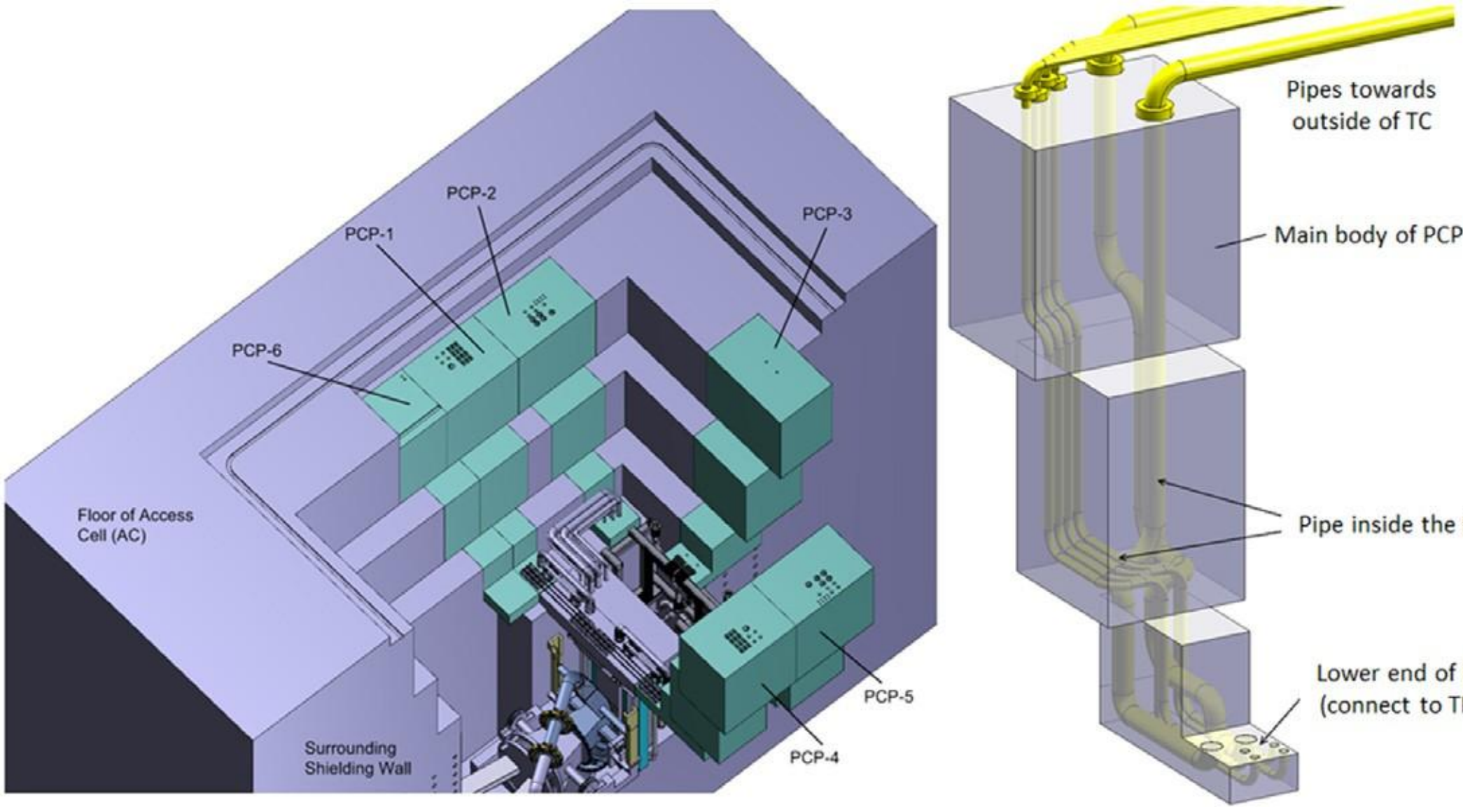

**Figure 16.** Left: the PCPs and the TMs (the USP and LSP are not shown). Right: preliminary piping inside PCP.

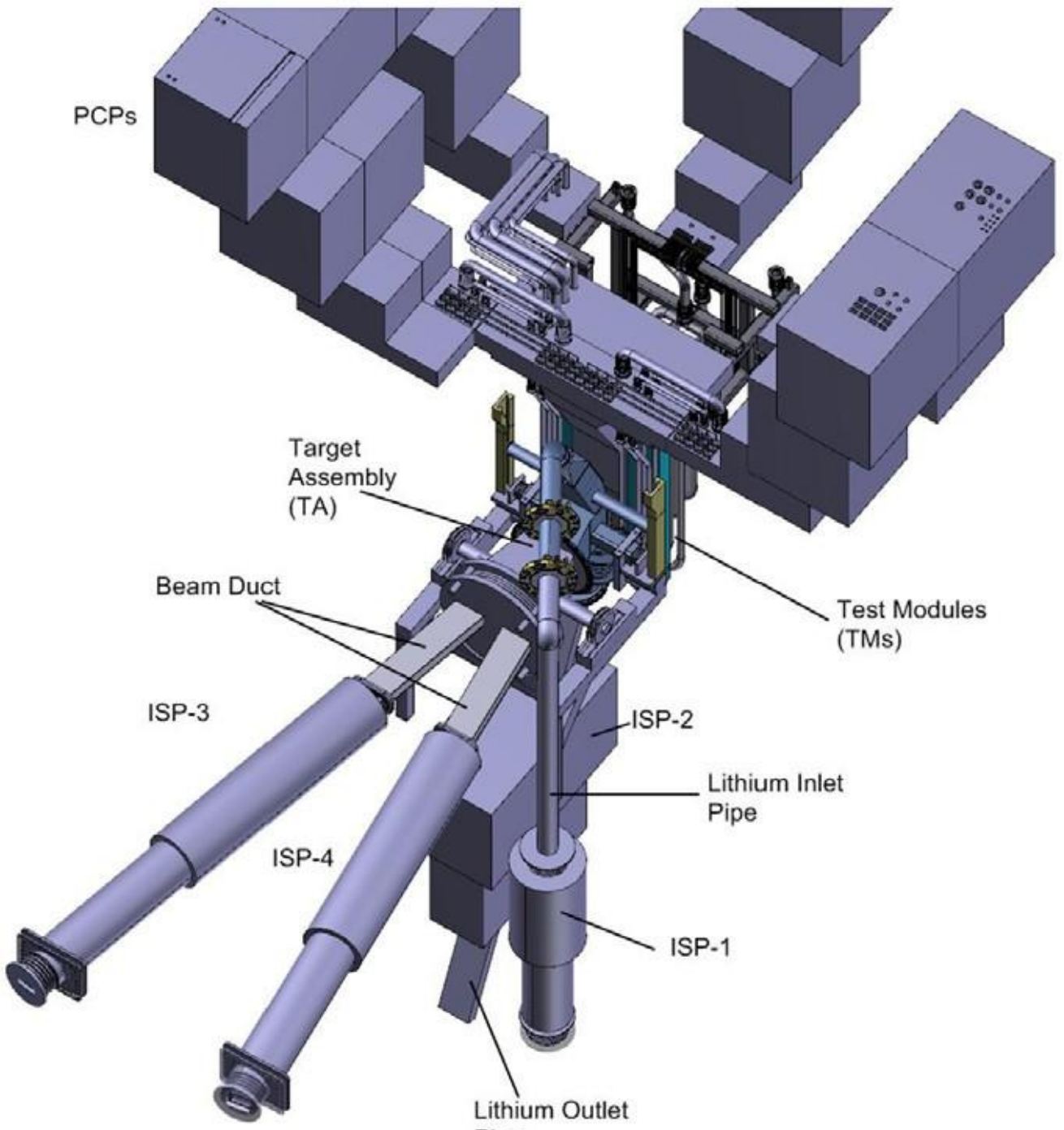

**Figure 17.** Arrangement of the ISPs.

The HFTM-V (see figure 19) is dedicated to the research on RAFM steels, to be tested in the temperature range 250–550 °C, with an option to provide irradiation up to 650 °C for oxide dispersion strengthened (ODS) steels [90]. The uncertainty of temperature for 80% of the specimens will be below ±3% thanks to an independent cooling of the capsules, which in addition count with an independent system of heaters, thermocouples and thermalisation of the specimens by filling the capsules with NaK-78 eutectic alloy. To measure and control the irradiation temperature, three up to six type-K thermocouples will be located inside the specimen stack.

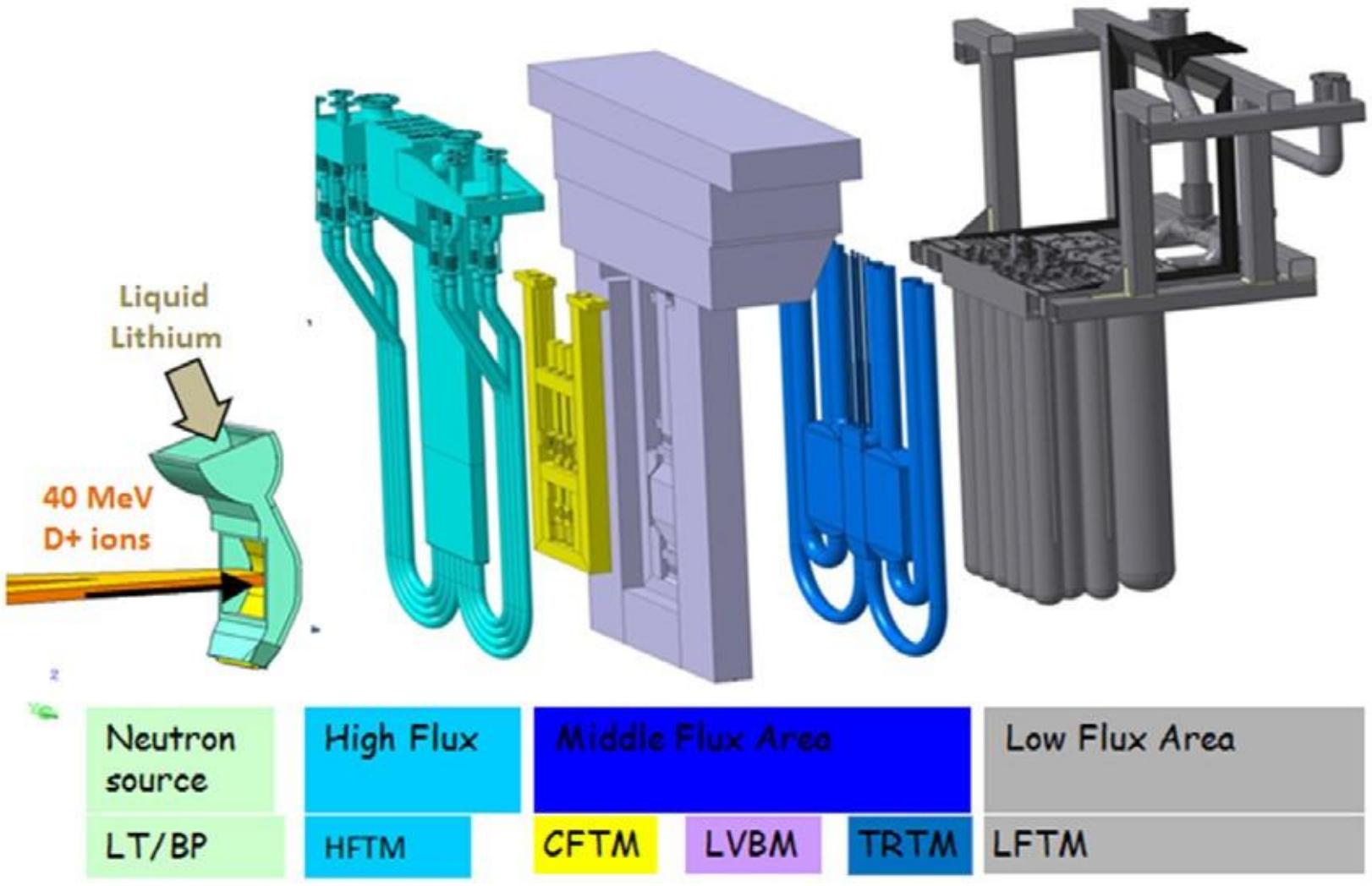

**Figure 18.** Target and TM arrangement in the TC—the neutron cloud and the TMs are shown symbolically.

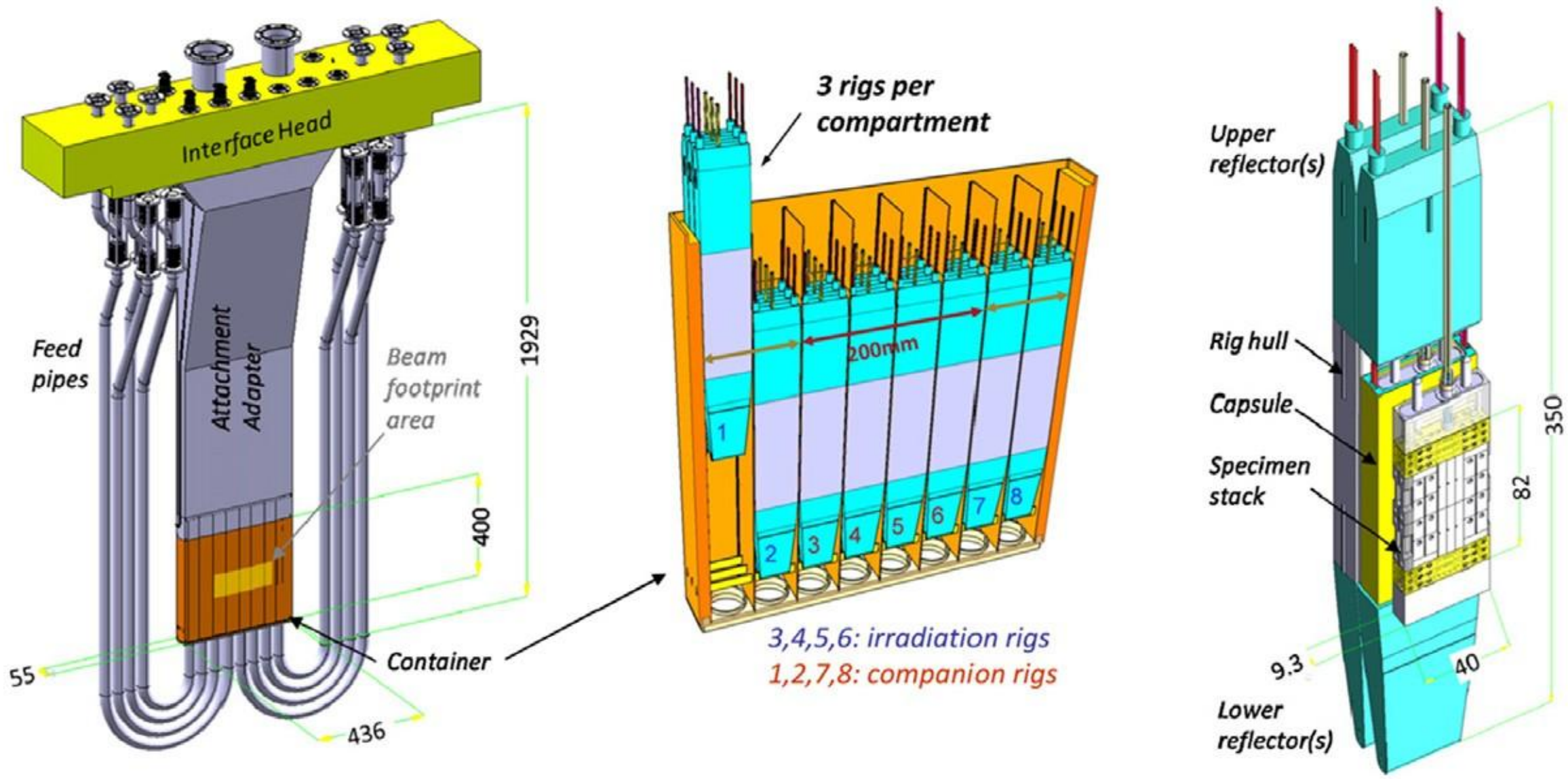

**Figure 19.** Design overview of the IFMIF HFTM showing assembly, compartments, irradiation rigs and capsules.

The thermocouple readings are the input to the control of the capsule's electric heaters. In addition, the specimens can be cooled from their temperature to below 200 °C within 15 min after the irradiation to avoid mitigation of the irradiation effects by annealing. The arrangement of the specimens in the HFTM is adapted to face the 200 mm$\times$50 mm beam footprint of the neutron source. The specimen positioning and dimensions of the reflectors are conceived to limit the neutron flux gradient to less than 10% of the individual sample's gage volume. The HFTM-V is built from a thin walled container divided into eight compartments, into which three rigs can be placed (a total of 8 $\times$ 3 rigs) (see figure 19). Specimens are arranged in the central 4 compartments that can house around 1000 specimens in a total of 4$\times$3 capsules where neutron flux gradients and flux levels are suitable for high quality irradiation experiments. The remaining four (2$\times$2) side compartments are also filled with rigs, but their function is mainly to act as lateral neutron reflectors and accommodate instrumentation, like fission chambers for online flux monitoring. It is to be noted that in these lateral compartments, the neutron flux amounts to only about 10% of the central positions, but the gradients are low, and can thus be attractive as additional irradiation space.

The HFTM interface head (TMIH) must provide media, energy and signal connections and mechanically fix the HFTM in the TC. The container is attached to the TMIH by a stiff attachment adapter. The stiff part of this adapter has a length of 1653.5 mm and a diverging cross section at the upper end for the TMIH. The dimensions of the HFTM attachment adapter fit in the available space between the TA and the medium flux test module (MFTM). The dimension of the wall thickness is 10 mm, and cylindrical bolts between the front and the back side provide additional stiffening against buckling due to the internal pressure. The attachment adapter also serves as a return duct for the helium cooling gas, and contains all the electrical connections (thermocouples, heaters, and n/gamma detectors) between the rigs and the TMIH. In turn, the helium feed pipes guide the coolant helium from the HFTM interface head down to the bottom of the container. There are eight pipes (one for each compartment), and on each side of the HFTM four pipes are arranged. On the top end, the pipes are fixed mechanically to the interface head. On the bottom end, they are welded to the bottom plate of the HFTM. To reduce the mechanical stresses coming from the differential thermal expansion of the pipes (cold) and the attachment adapter and container (hot), the pipes are equipped with expansion bellows.

The HFTM-H is designed for the irradiation campaign under high temperature conditions, up to 1100 °C for $SiC/SiC_f$ composites and refractory materials, such as tungsten alloy [91]. It is considered as an alternative for the HFTM-V in different irradiation campaigns and it will be arranged at the same position as the HFTM-V. In the beam footprint region, the HFTM-H is built from a thin walled container divided into three arrays of capsules, into which three capsules can be placed each (a total of 3×3 capsules). The array of capsules is separated by narrow coolant channel (1 mm × 200 mm) flowing helium gas at up to 0.5 MPa. The helium flow is introduced at the bottom of the HFTM-H container through the connecting pipe and passes the flow straightener made by the porous material and manifold before entering each cooling channel. For the irradiation at high temperatures, the gap between the capsule walls and the specimens is filled with helium gas at the same pressure as in the coolant channel. The container also integrates neutron reflectors.

The CFTM consists of three parallel testing machines, mounted on a frame, which will be operated independently [92] and allows for the testing of creep-fatigue behaviour under irradiation conditions. It will be installed in the first raw of the medium flux area of the TC and will be exposed to intense radiation fields, including both neutron and gamma fields. The heat absorbed by the different parts of the module will be extracted by means of a helium gas (coolant) passing through independent cooling channels. Elements required to control the specimen temperature at the targeted value will also be included. The testing machines will include radiation sensitive components, such as load cells and extensometers, and therefore they must be chosen with adequate radiation hardness.

The TRTM is a system that contains specimens of the solid breeders (lithium orthosilicate or metatitanate) or beryllium based materials in controlled conditions, with connections to a purge gas delivery system and a measurement system, to measure the tritium release online during irradiation [93]. It consists of three main components: (1) the containers; (2) the coolant and the purge gas tubes; and (3) the TMIH. The TRTM has three containers (a middle container and two duplicate lateral containers). The middle container encloses eight rigs arranged in two rows (2×4) while each lateral container encloses a lateral neutron reflector that consists of 12 graphite bars. Each rig contains one capsule that is packed with pebbles of the tested material forming a pebble bed. At the capsule top and bottom, upper and lower filters are used to constrain the pebbles whilst allowing the purge gas to flow. The purge gas is fed through the inlet tube (along the capsule axis) from the top to the bottom, and then it flows upward throughout the pebble bed and exits through the outlet tube. Coolant gas flows to the container in four inlet tubes and exits in four outlet tubes above the containers to remove the nuclear heat.

The TRTM system extends up to TMIH, which is the interface system between the TC and the TRTM. All of the cables and pipes from the TRTM penetrate this TMIH and then are further connected to the ancillary systems. The irradiation capsule has a cylindrical shape with both ends having a filter to constrain the pebbles of the tested material. It is also surrounded by a cylinder (heater-can) which hosts the heater. The gap between the capsule and the heater-can is filled with stagnant purge gas to make a temperature gradient between the capsule and the heater-can, and to avoid any differential pressure across the capsule wall. Each capsule has its own inlet/outlet purge gas tubes and is equipped with thermocouples to measure its internal temperature distribution.

The LBVM is designed to perform irradiation tests related to liquid breeders for future fusion reactors [94]. The LBVM will be focused, as the first approach, on experiments related to the short term blanket concept based on LiPb as a breeder and He as cooling. The present configuration of the LBVM consists basically of a stainless steel 316LN container capable of housing 16 rigs. Each experimental rig will support one experimental capsule containing LiPb, specimens and the associated instrumentation. The capsules are closed recipients that contain the liquid breeder. These capsules will be installed inside cylindrical rigs that, in turn, are installed inside the LBVM container. Each capsule will also be equipped with electrical heaters in order to get the required operational temperature and the required instrumentation for the experiments. The capsules contain a thermal insulation to avoid heat transmission through the capsule support to the rig walls. There are 16 EUROFER cylindrical capsules in the current design of the module. Each capsule will be dedicated to a particular experiment.

The rig is the recipient where the capsule is housed. The central part of the rig has the same cylindrical shape as the capsule, but a narrow gap (2 mm) between the rig and the capsule allows for the circulation of the purge gas. The lower part of each rig is a small diameter tube which is welded to the bottom part of the container and also to the entering purge gas line. The upper part of each rig is also a small diameter tube which is welded to the upper part of the container and also to the exit purge gas line. It has a rectangular section in most of the zones, except in the testing area where 16 cylindrical compartments are built to house the rigs. The upper and lower pipes of the rigs are welded to the upper and lower part of the container. The container also serves as a common collector for the returning He cooling gas.

LBVM will be equipped with a set of instrumentation. Some of the diagnostics will be used for collecting information about the experimental conditions, e.g. temperatures, and some others will be directly related to the safety aspects of the module.

The NSS are to be located in the MFT region to optimise the irradiation conditions by moderating the neutron spectrum to simulate the conditions of the breeder zone in the DEMO reactor [94]. NSS materials shall withstand significant radiation and thermal loads. Tungsten was selected as a spectral shifter material due to its high melting temperature, good thermal conductivity as well as good neutronic properties such as high scattering ability. The NSS consists mainly of a hollow structure of 316LN aiming to house eight tungsten plates; this container presents two inlet pipes and two outlet pipes through which the cooling gas circulates. In addition, there are two guiding plates on both sides and a TMIH on the top for the purpose of fixing the module to the TC.

The LFTM [88] can be used for multiple irradiation campaigns because of lower neutron damage accumulation in the material. Consequently, the LFTM is made up of two parts: (1) a cooling matrix which will be permanently installed in the TC; and (2) experimental rigs which could be inserted during each irradiation cycle if required. The cooling matrix provides a helium flow around the experiment rig. The helium cooling pipes of the cooling matrix will be permanently connected to the PCP. The only connection that should be made for an actual experiment is the experiment wiring and/or extra piping for a variable atmosphere, etc. This design clearly reduces the RH operations since, for standard experiments, no additional piping needs to be connected. Also the radioactive waste is reduced to a minimum. The cooling matrix operates at a constant pressure difference regime which allows an independent control of the helium flow in each rig. The helium flow around each experiment can be adapted individually by means of a flow restrictor inside the experimental rig. The connection of the experiment head and the cooling matrix needs to be leak-tight to avoid helium leaking to the TC. This sealing can be created if necessary by use of a metal axial compression seal. This type of sealing is radiation tolerant and allows high pressure and thermal cycling.

The STUMM presents the following objectives: (1) the characterisation of the whole radiation field in the TC area during the commissioning phase; (2) the measurement of the spatial distribution of neutrons and photons; (3) the characterisation of the temporal evolution of the radiation field; (4) the check of the correct functioning of the foreseen instrumentation; and (5) the validation of neutronic calculations and models used for the engineering design of the TMs [96]. Two different STUMM concepts have been assessed: (1) the duplication of the HFTM-V design but filling the modules with a large amount of instrumentations; and (2) the LFTM-like with space for a diversity of instrumentation and placed adjacent to the HFTM-V-like type.

Under the validation activities, a prototype of the HFTM-V has been installed in the HELOKA-LP helium loop facility, where it has been operated under IFMIF design operation conditions, in what regards mass flow, gas pressure, gas temperature and electrical heating of the capsules [97]. The test objectives included the verification of the temperature control strategy, assessment of flow induced dynamic loads on the rigs, the attachment structure and the helium pipes, and the definition of operational modes. In turn, three capsules, filled with specimens and fully instrumented, have been tested under irradiation conditions in an experimental reactor [98]. Obviously, considerably lower total doses than anticipated for IFMIF end-of-life have accumulated in this campaign; the main objectives were to learn the stability of the heaters and the capsule temperature control under irradiation conditions.

Due to the limited irradiation volume available in the HFTM, small specimens have been successfully developed under the IFMIF/EVEDA frame [99, 100] profiting from the wide experience on small specimens related to material developments for fission reactors [103]. Though the standardisation is not accomplished, what will likely be indispensable for a fusion power plant licensing, the shape of the specimens have been defined allowing up to 90 specimens per capsule (~1000 specimens per irradiation campaign) for the full characterisation of 2 sets of materials per capsule (~40 specimens are required for a full characterisation) (see figure 20).

In addition, tests in a hot cell have been successfully carried out with specific developed tooling to demonstrate the feasibility of assembling and disassembling of the capsules.

*4.4.3. The AC and test modules handling cells.* The AC is a closed cavity which is located directly at the top of the TC and is designed to fulfil the following requirements: (1) to provide sufficient accommodation space for RH tools for TM and TA maintenance operations inside the TC; (2) to provide sufficient space for RH tools to perform operations inside the AC; (3) to ensure sufficient shielding between the AC and the non-irradiation areas when the TC is open; (4) to provide sufficient lodging places for removable shielding materials from the TC; (5) to provide convenient access of RH tools to the inside of the TC; (6) to provide a reliable and safe transferring pathway for all of the TMs, TA, and other removable components (including the steel framework of the double liner); and (7) to provide sufficient shielding to the surrounding areas when the inner of the AC is exposed to intense radiations.

The TMHCs are a series of hot cells that are located functionally between the AC and the PIEF. The TMHCs are designed to fulfil the following requirements: (1) to provide sufficient accommodation for the RH tools for handling TMs and other components that are transferred from the AC (TMs, backwall plate, TA); (2) to provide sufficient space for the RH tools to perform operations on the TMs and other components; (3) to allow TMs to be disassembled and reassembled using RH tools; (4) to ensure sufficient shielding between the TMHC and non-irradiation areas; (5) to provide convenient and safe access to AC, and other adjacent areas; (6) to separate different areas according to radiation and contamination levels.

The AC is located directly above the TC and it is the room where the main RH systems are installed. The AC is the only access to the TC and also contains the storage areas for the top SPs, metal sealing plate, connection pipes, all tools for the RH equipment, etc. The floor of the AC also covers the top areas of the test facility ancillary system (TFAS (cooling, power supply, control etc)), helium room, tritium room etc. Furthermore, the AC reaches over all of the TMHCs, including their maintenance and utility rooms.

RH devices and robotics are considered as the standard maintenance tools inside the AC because of the activation and contamination of components exposed to high radiation. Because handling of the activated TMs and TA is required inside the AC, the thickness of the walls all around the AC is presently estimated to be about 1.5 m of concrete as BS.

During the irradiation procedures, the AC is accessible for operators; during the maintenance periods, the AC is accessible only if the TC and the TMHCs are closed and no activated components are present in the AC. The TC can be opened to the AC above at beam shut off. Handling robots and cranes move modules and targets for replacement from the TC position to the handling cells. The thickness of the BS is designed to allow access of the AC even during irradiation periods. Inert atmosphere inside the AC is not required.

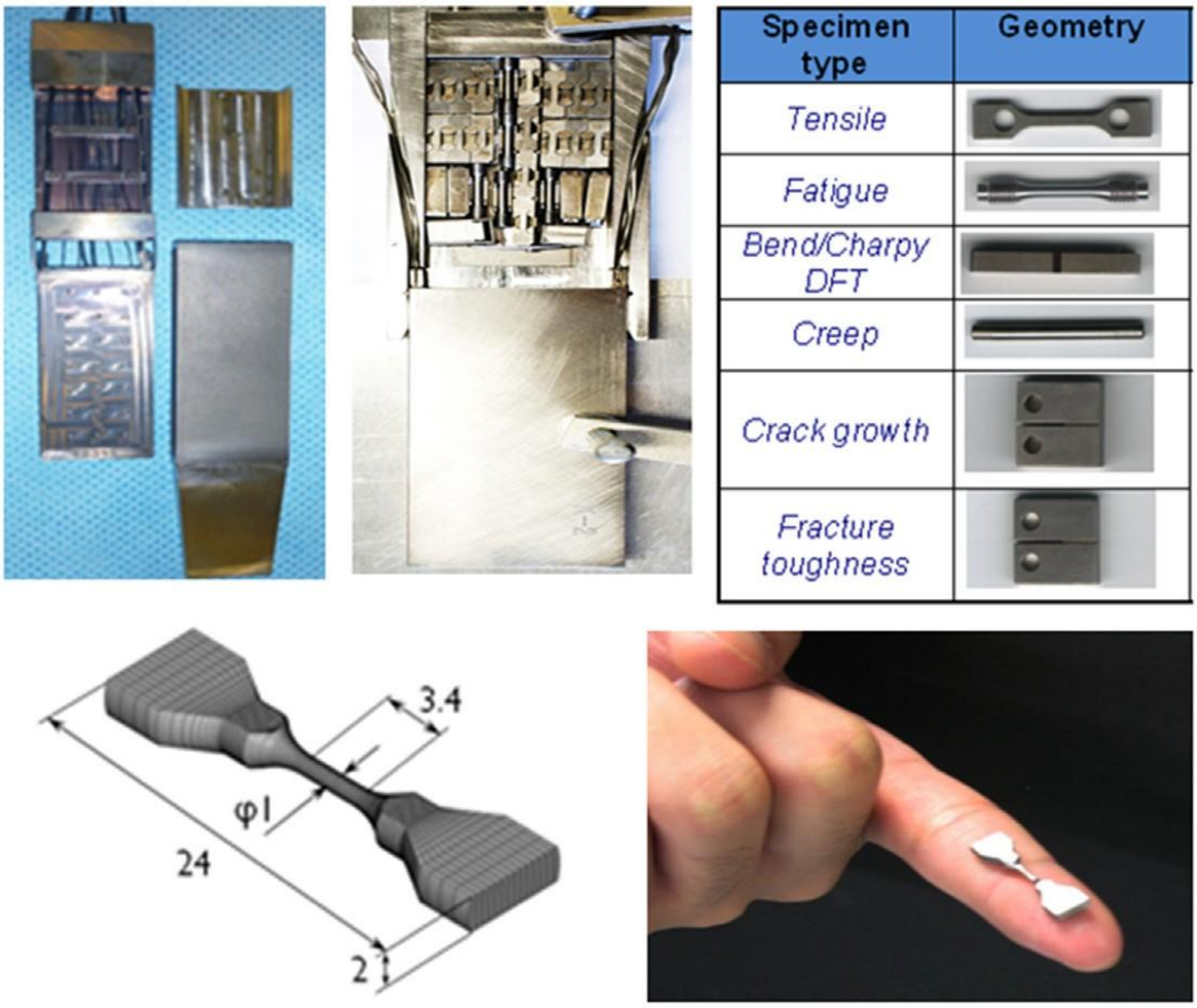

**Figure 20.** Rig, capsule filled with small specimens, shape of small specimens housed (2 × 45) in each capsule for a full characterisation of one set of materials and dimensions of tensile specimen.

*4.4.4. RH system.* The RH system for the TF [101, 102] is planned to be installed in the AC and the TMHCs to perform operations on the components in the TC and the TMHCs (the installation and extraction of the TMs, transporting components between TC and TMHCs, mechanical processes on the irradiated materials in TMHCs, etc). The design requirements for the RH system for TF include: (1) capability for all the required RH operations and procedures; (2) fulfilment of IFMIF availability; (3) recovery and rescue operations; (4) integration with other IFMIF systems, in particular the RH System for the target facilities; (5) flexibility to cope with potential upgrades of the TF.

The RH system includes two cranes in the AC: (1) a heavy rope overhead crane (HROC) to lift and transfer any weighty component in the AC requiring moderate positioning accuracy (TC cover, TC shield plugs, PCPs, etc), capable of handling 120 t and will present an auxiliary hoist (15 t load capability) to help the lifting operations of the most irregular pieces; and (2) an AC mast crane to support and locate the servo-manipulator and lift and slightly tilt the light weight components (<1 t). It is assumed that the middle weight components (<3 t) could be moved by the hoist attached to the HROC.

*4.4.5. Ancillary systems.* The TFAS host the dedicated equipment to supply sub-systems of the TFs. The TFAS supply energy, heat sinks (through helium gas or water flows), media (purge gas flows) and control infrastructure to the TMs, the TC, and other client systems. On the other hand, the TFAS receive their energy and media from the CFs.

### *4.5. The post-irradiation and examination facility*

The PIEF will mainly perform PIE of the irradiated specimens in order to generate a material database [104] (see figure 21). PIEF will also perform some PIE after *in-situ* tests. The PIEF will provide the capability to conduct mechanical properties and the other properties on irradiated materials, and it will have also the ability to characterise the fracture surfaces after the failure test. It will also have the capability for long term storage of irradiated material for further future analysis. As a main assumption of the functional definition, PIEF must be able to perform in 1 year the PIE of all specimens for 4 of the rigs set in the HFTM, and in 3 years all tests of the specimens for all 12-rigs of HFTM. As a reference, the list description and capabilities of other PIEFs all over the world, summarised by IAEA, has been used [105].

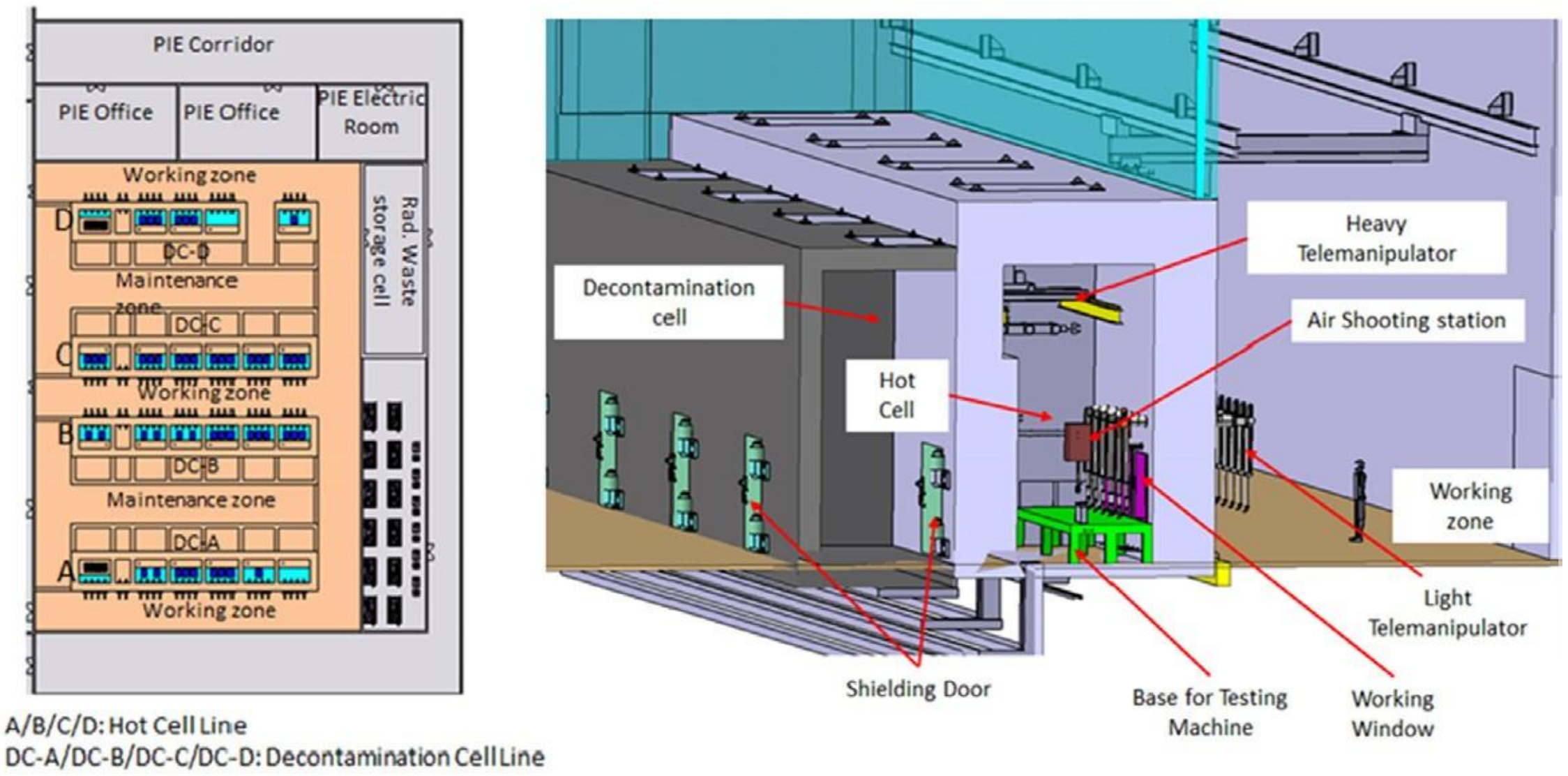

**Figure 21.** Left: two-dimensional layout of the PIE facility. Right: isometric view of a line of the hot cell laboratory.

The PIEF is placed in a wing of the main building in order to minimise the handling operations of irradiated specimens. It will not only allow for the testing of irradiated specimens exposed to the three irradiation levels but also their metallographical characterisation after destructive testing [106].

### 4.6. The CFs

The design of the CF, comprising buildings, site infrastructures and plant services, has been carried out with the support of engineering services of specialised industry in Japan and Europe [107]. The layout and the corresponding 3D models were developed based upon a comprehensive analysis of the functions and implantation of the different rooms, the description of the whole plant and of each room (including materials flow, access routes, handling, lift, etc), as well as the main equipment footprints (volume/space reservation) and routing plans of the main heating, ventilation and air-conditioning (HVAC) ducts, pipes and cable trays. The objective was to allow the management of the IFMIF plant 3D models from the onset and all along the design process by continuously cross checking the clearance and potential interferences to eventually allow for a complete integrated model of the IFMIF plant and the different systems/facilities inside the building.

The breakdown of the IFMIF plant services includes: (1) the HVAC system (both industrial and nuclear); (2) the heat rejection system; (3) the electrical power system; (4) the service water and service gas system; (5) the radiation waste treatment system (including both solid and liquid waste as well as a complex exhaust gas detritition system); (6) the fire protection system; (7) the access and security control system; and (8) the radiation monitoring system. The design of each system was developed progressively, firstly by establishing a sound design basis starting from a system functional description, followed by the identification of the corresponding interfacing systems and the technical requirements imposed by them, and ending with the definition of the process flow diagrams and basic equipment layout. Once the technical requirements were identified and the design basis established, the system designs were further developed. Piping and instrumentation diagrams, key-one line diagrams, and equipment lists for the different systems, as well as a layout plan of the main equipment and the routing of ducts, piping and cable trays were defined and eventually integrated into the 3D model of the building.

## 5. Safety analysis

IFMIF safety objectives are related to safety principles coming from the IAEA safety standards Series No Sf-1, IAEA safety assessment for facilities and activities (IAEA GSR No 4) and Basic Safety Standards for Protection Ionizing Radiation and for Safety of radiation Sources (IAEA safety Series No 115). Thus, in the IFMIF safety approach, the priority is to prevent accidents by a robust design to control hazards and to mitigate consequences of postulated initiating events. It means that safety principle 'Defence in depth' has been integrated during engineering design, to ensure each system will meet safety objectives in a coherent way.

A safety analysis has been performed individually on each IFMIF facility, except for PIEF that has been partially analysed as part of the CF. This study has not only covered radiological safety but also conventional safety, including: fire, chemical (including liquid metals), cryogenic, pressure and vacuum, anoxia, electrical and magnetic fields, mechanical and human factors. Safety functions (see table 3) to prevent or to mitigate against radiological hazards are: (1) the confinement of radioactive material by confinement barriers and associated confinement systems; and (2) the limitation of external exposure to ionising radiation.

In the AF, the release of radioactivity under normal operation conditions is considered of low relevance. However, activation is a limiting factor for the accelerator components for hands-on maintenance; the choice of low activation structural materials and materials for beam line elements are mandatory to optimise maintenance and waiting periods after shutdown. Calculations on air activation (after 11 months of continuous irradiation) have shown that $^{41}Ar$ is the critical radionuclide to take into account for beam transport lines rooms (around 2.4 h are needed for $^{41}Ar$ decay to breathing levels); $^{16}N$ > 1.0 DAC (derived air concentration), is also produced but its activity decreases in a few seconds to unrestricted release of activated air. A tritium hazard is a concern essentially to air activation in the beam line and cooling water system. Airborne tritium after 11 months of continuous irradiation reach 2.3 MBq (<1 DEL). In any case, air renewal of vault atmosphere (HVAC system) should be performed in advance, before workers enter for maintenance. A radiological hazard associated with an activated corrosion product (ACP) is a concern within the cooling system driven by the emitters $\beta$, $\beta\gamma$ with the main ones being $^{51}Cr$, $^{54}Mn$, $^{56}Co$, $^{57}Co$, $^{60}Co$, $^{55}Fe$, $^{60}Co$ and $^{64}Cu$.

**Table 3.** Safety functions for IFMIF plant.

| | Safety function | | Detailed safety functions |
|---|---|---|---|
| 1 | Confinement of radioactivity | 1(a) | Process confinement barrier |
| | | 1(b) | Building confinement barriers including systems for maintaining depression and filtering/tritium clean-up effluents |
| 2 | Limitation of exposure | 2(a) | Shielding to limit exposure and ALARA (as low as (is) reasonably achievable) principle |
| | | 2(b) | Access control |
| | Service functions | | Detailed service functions |
| 3 | Protection of systems for confinement and limiting exposure | | Fire detection/mitigation |
| 4 | Supporting functions | 4(a) | Providing auxiliaries essential for implementing safety functions (electrical power supply, instrumentation and control), compressed air, etc) |
| | | 4(b) | Monitoring plant status: safety functions, radiation monitoring |

In the LF, the backwall plate is the most heavily exposed component to the high neutron irradiation flux produced in the flowing lithium (an induced activity of $2 \times 10^{18}$ Bq m$^{-3}$ 1 month after shutdown is estimated). The radioactive materials to be considered are: (1) tritium inside a lithium loop (the maximum inventory in the lithium loop is 6.7 g per year of tritium if traps are replaced annually; (2) activation of corrosion products in liquid lithium; (3) $^{7}Be$ (its production will be around 1.5 g per year, mainly removed in the traps together with tritium) with maximum $^{7}Be$ inventory of 0.31 g per year if traps are replaced annually); (4) activation product in the components considered as solid wastes; and (5) activation and activation corrosion products in the liquid wastes. Additionally, lithium fire and anoxia, due to argon atmosphere in some of the rooms, could become more important hazards than tritium.

In the TF, the TC is obviously where the highest quantity of radioactive materials will be generated. The TC itself is designed with sodium free concrete and a 316 L stainless steel liner. Tritium generation in concrete will be negligible. An extensive safety neutronic study has been performed for the different TMs (HFTMs, LBVM, TRTM) to evaluate radionuclides production as a consequence of one year of irradiation. The analysis shows that the main risks come from: (1) tritium created in lithium ceramic breeders and LiPb specimens, as well as in EUROFER and stainless steel irradiation; (2) $^{55}Fe$, $^{54}Mn$, $^{51}Cr$, $^{60}Co$ from EUROFER and stainless steel irradiation; (3) activated corrosion products and activation products in cooling systems (tritium, $^{210}Po$ from PbLi (from LBVM), $^{22}Na$ and $^{39}Ar$ from NaK liquid metal activation (from capsules of HFTM), impurities activation and $^{14}C$ from graphite activation (spectrum shifter, reflectors)).

In the CFs, the radioactive materials to be considered are mainly tritium, activation corrosion products such as aerosols ($^{7}Be$, $^{54}Mn$, $^{60}Co$ etc), activation and activation corrosion products in the components considered as solid waste and in the liquid waste.

During operation of the accelerators, provisions to prevent radiation leakage to the upper and lower floors are implemented in the design. There are no direct penetrations to the upper floor in the radiation isolation rooms (RIRs) or beam transport room (BTR), but there are penetrations for the RF guides in the ceiling of next piping and cabling penetration spaces, thus the contribution to doses at the upper floor is negligible. The maximum values obtained are due to neutrons scattered from the lithium target, which produced photons in the ceiling concrete. The highest recorded dose rates are below 0.5 $\mu$Sv h$^{-1}$, which match the occupational objective. The contribution of beam dump to the dose rates in the upper floor are mitigated with up to 2.3 m width on the concrete in the floor. A polyethylene thickness of around 65 cm is needed in order to achieve the dose required in adjacent rooms. Dose rate maps during AF operation are shown in figure 22 indicating the origin of the different main contributions.

In addition to the accelerator components, the prompt dose due to deuterons and neutrons also has an impact on the TC and LF components through their activation. The TC floor is 2 m thick to provide sufficient shielding to the QT, which is located in the room below. However, the QT is still activated during the irradiation due to intense neutron streaming through the lithium outlet pipe and diffused neutrons. The most activated components around QT are the heater layers of the outlet channel due to neutrons streaming along the channel. Gamma rays emitted inside the concrete wall are shielded partially, but the contribution from those emitted outside the wall is still very significant. The major nuclides contributing to the dose are $^{56}$Mn ($T_{1/2}$ = 2.6 h), $^{58}$Co ($T_{1/2}$ = 71 d) and $^{54}$Mn ($T_{1/2}$ = 312 d). After the decaying of $^{56}$Mn, the dose level becomes almost constant for around 100 d.

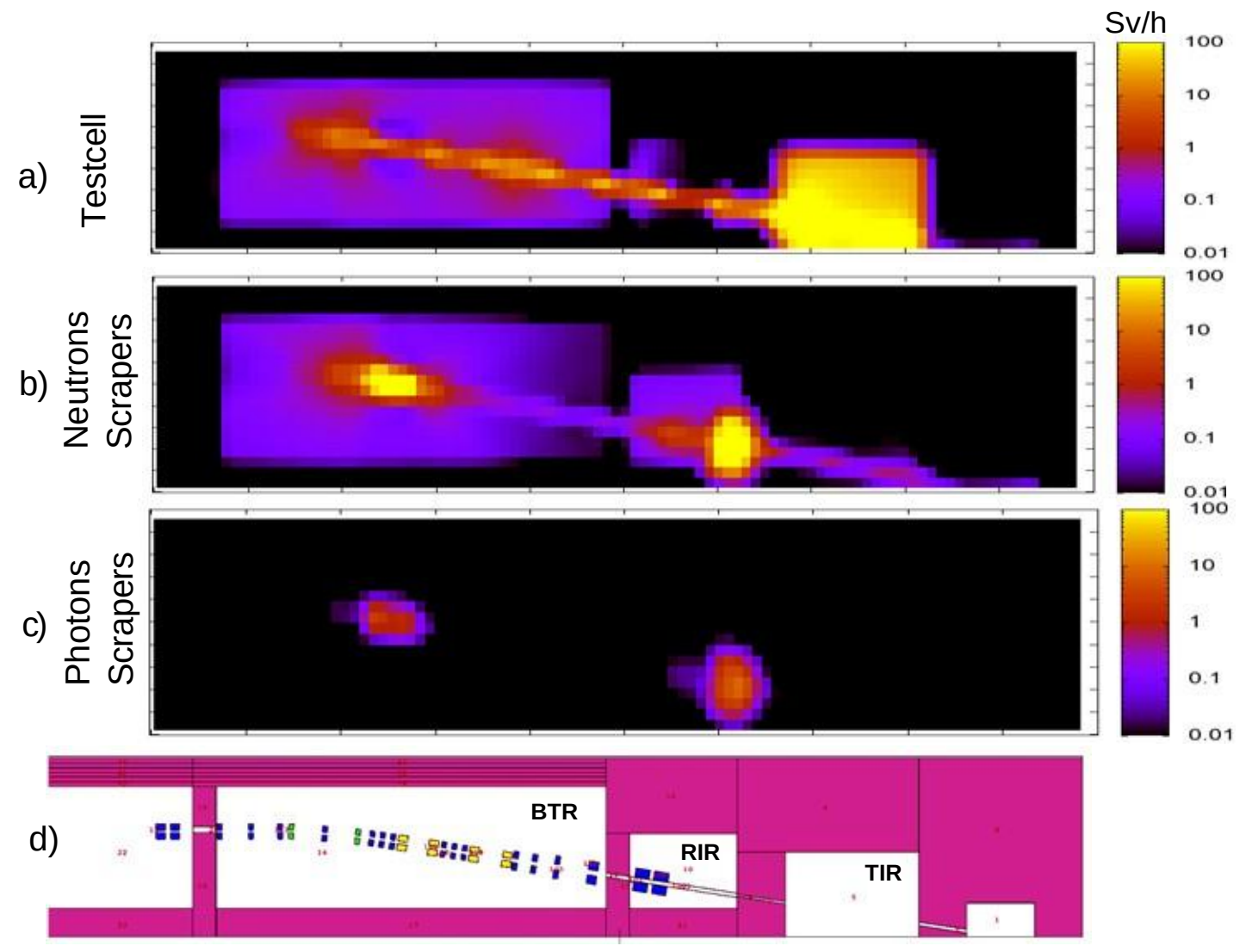

**Figure 22.** Effective dose inside BTR, RIR and TIR during operation (Sv h$^{-1}$): (1) contribution from neutrons coming from the LF; (2) contribution from the neutrons generated in the shielded scrapers; (3) contribution from the gammas generated in the shielded scrapers; and (4) the scheme of the rooms involved.

The biological dose rate expected in the AC on top of the TC USP is less than 100 $\mu$Sv h$^{-1}$ assuming present design parameters with a total concrete thickness of 250 cm. A calculated dose map is shown in figure 23. Although the dose rate exceeds partly 25 $\mu$Sv h$^{-1}$, limited access to the AC during operation is possible for workers. The TC BS is designed to be permanent and is integrated in the IFMIF building. It is assumed that during the complete IFMIF life span, the BS of the TC can withstand the neutron and gamma irradiations and provide sufficient shielding to adjacent rooms and cells.

While the injector presents no activation after 1 h cooling time, the residual contact dose rate for RFQ, after 1 h cooling time, is dominated by short-lived $^{64}$Cu (12.3 h). A waiting time of 1.5 d is required before complying with dose rate requirement at 100 cm from the equipment, reaching a steady value of 5 $\mu$Sv h$^{-1}$ after 3 d cooling time. For periodic maintenance, no problem is foreseen for hands-on operations in the whole RFQ, although portable shielding could be needed. In the SRF, residual doses around the last cryomodule are 7.4 times higher than around the first one; the dose rate after 1 h cooling time in the former is around 50 $\mu$Sv h$^{-1}$. With these values, an estimated total cooling time of 4 h is needed in order to reach hands-on maintenance levels. In the case of HEBT, residual dose rates have been calculated from the radioactive inventory of scrapers, magnets and concrete walls. The radioactive inventory in magnets was calculated considering their dimensions, average neutron flux and homogeneous composition mixture of iron and copper. Comparisons with the use of aluminium showed higher delayed doses after 1 d cooling but lower after 1 week cooling. The results show that residual doses from activated scrapers are important, but effectively contained inside the scraper shield of the BTR. The iron shield in BTR has a contact dose rate of 16 $\mu$Sv h$^{-1}$ after 1 h cooling going down to 1 $\mu$Sv h$^{-1}$ after 1 d. The activation of magnets is also important in the RIR, producing residual doses much higher than the projected dose rate target for unrestricted access. The radiation from scrapers is mostly contained inside the shields. On the contrary, contact doses from unshielded magnets are up to 150 $\mu$Sv h$^{-1}$ after 1 h cooling, being reduced to 31 $\mu$Sv h$^{-1}$ after 1 d cooling. Then, magnets have to be shielded in order to make the room accessible.

During maintenance operations, the most limiting situation occurs when one accelerator is operating at 100% and the other is under maintenance. In this case there are four categories of sources to be considered: (1) particle streaming from the lithium target, due to interactions with the other beam; (2) particles crossing the wall between collimator and scrapers previously activated by deuterons; (3) magnets and concrete activated by neutrons; and (4) beam losses in the accelerators. The third and fourth contributions are initially dominated by short-lived radioisotopes, and thus decay in time after shutdown. The first and second contributions are constant in time. A triple shutter (down- to upstream 20 cm steel, 30 cm polyethylene, 10 cm lead) is proposed to assure that maintenance operations are feasible in one accelerator when the other one is running by particles streaming and a 1.5 m thick concrete separating wall between the accelerators is present. The IFMIF shielding approach has been performed based on the prompt doses and residual doses calculations summarised above.

A hazards evaluation has been performed through failure mode, effects analysis and failure mode, effects and criticality analysis (FMECA) techniques applied to each facility individually. This approach has been used to identify possible failure modes, their causes and effects and in case of FMECA, to consider also criticality as a ranking of severity of failure modes to allow mitigation. This has allowed for the anticipation of failures and the implementation of a design focusing on the mitigation of consequences, thus an increase in safety of the analysed system. A risk matrix was developed as a relative measure of the consequences and likelihood of an unsafe situation and the severity of the event. This risk matrix can be used to prioritise actions to mitigate a potential incident/accident and to reduce the weakness of the components from a safety point of view. Related to the risk matrix, a quantitative approach has been defined in terms of the number of barriers or safeguards, being the technical barriers and operational barriers related to the risk ranking categorisation obtained from FMECA analysis. From this evaluation, the safety requirements for the different systems have been reviewed and completed and potential initiating events have been identified to be followed during future construction phases and procedures development.

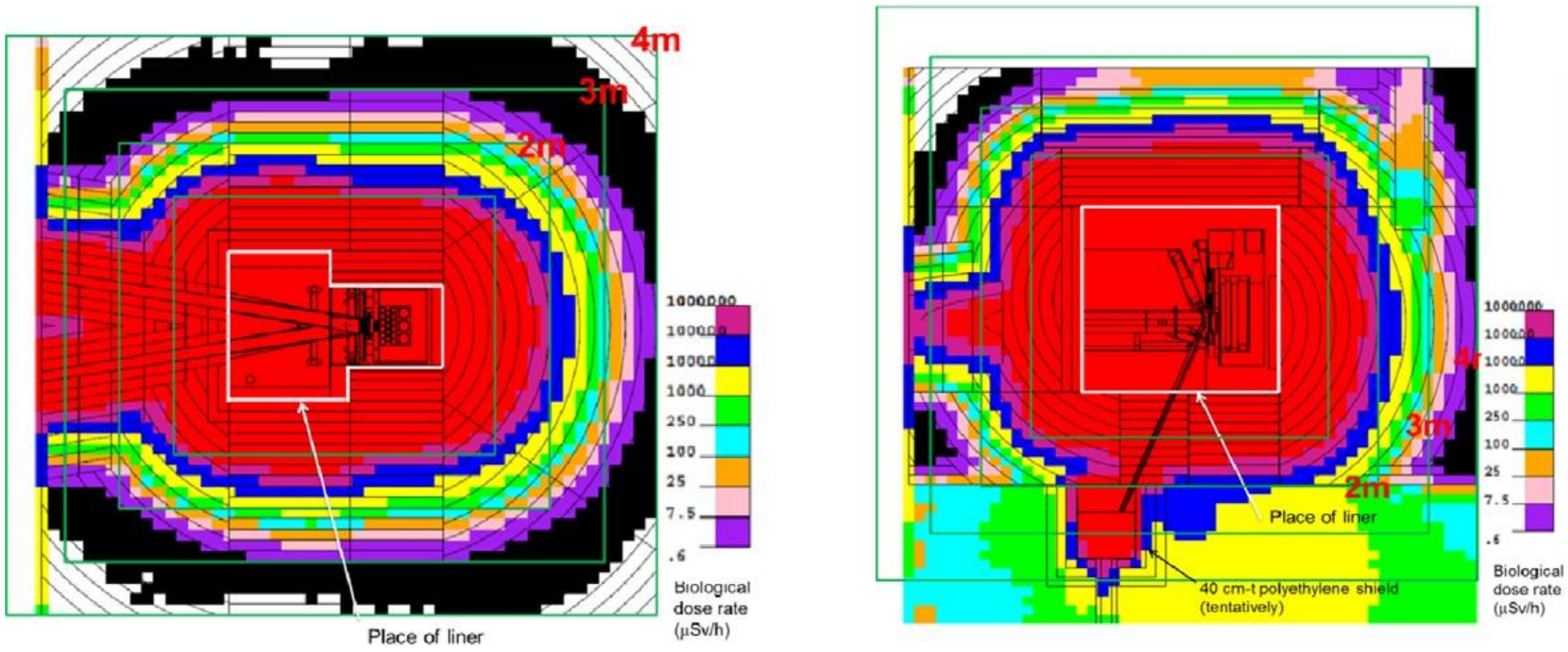

**Figure 23.** Left: dose map of TC during irradiation ($\mu$Sv h$^{-1}$), horizontal cut at beam level. Right: dose map of TC during irradiation ($\mu$Sv h$^{-1}$), vertical cut at target centre.

IFMIF safety implementation is based on the 'defence in depth' principle. Safety referential documents for IFMIF engineering safety implementation have been prepared. Safety important class system structures and components are established towards licensing fulfilment. SIC-1 are those required to bring to or to maintain the IFMIF plant in a safe state; SIC-2 those used to prevent, detect or mitigate incidents or accidents; SR those systems and components identified as safety relevant but are not SIC-1 or SIC-2. Systems such as the lithium circulation system, personnel protection system, HVAC outlet, exhaust gas processing, primary cooling systems, tritium laboratory or fast beam shut down are some of the systems classified as SIC-1.

## 6. The availability of IFMIF: RAMI and maintenance studies

The availability goal for IFMIF is 70% over the calendar year, which together with its specifications regarding damage rate in iron (>20 dpa fpy$^{-1}$ in the high flux region) is directly linked to the main mission of IFMIF. The irradiation cycle is established in 11 months, mainly based on the lifetime expectations for the TA. This is broken down in one long maintenance period of 20 d for general maintenance (mainly in the TA and TMs replacement) and long term accelerator maintenance, and another intermediate maintenance period of 3 d for short-term maintenance activities in the accelerator and other ancillary and conventional systems.

RAMI analyses have been performed in order to identify critical components and to develop strategies to reduce downtimes and increase reliability. They have assessed the components design and they have allocated the desired availability to each sub-system so that the high RAMI requirements have been met successfully. Detailed analysis of the different facilities have led to high reliability and maintainability design evolution, and have brought up design proposals that, once implemented, have shown promise of becoming a way towards the IFMIF goal.

The methodology has been adapted to each particular case. The general methodology included FMECA, fault tree analysis and reliability block diagram analysis. The facility availability requirements have been allocated among systems and sub-systems and each system has been analysed in order to demonstrate the compliance with the goal or quantify the deviation from it. Design recommendations and provisions have been defined as RAMI strategies to approach the availability goals. Importance and sensitivity analyses have been carried out in order to detect critical parts and parameters that affect IFMIF availability. According to these results, a specific focused analysis has been developed for different facilities or systems depending on their RAMI characteristics. In that sense, for instance, the TC is characterised for the huge mean down time needed every time a failure happens. This leads us to focus on the analysis, on the one hand, of the reliability of components inside the TC and degraded operation of it and, on the other hand, on the reduction of access times for maintenance activities. However, for the case of the AF, the operation will be more discontinuous due to the unavoidable trips and short stops of this kind of facility, therefore specific approaches to estimate the influence of such operation patterns on availability were needed. The RAMI database has been generated for the analysis mainly through the collection of reliability data from related field facilities [60, 108, 109] (accelerator facilities, fusion, fission fields, generic databases, etc), and also through estimations of the mean down time of the different components (cooling time, detection time, access time, repair time, recovery and tuning time, etc).

**Table 4.** Inherent availability requirements for IFMIF facilities.

| IFMIF facilities (and systems) | Inherent availability (%) |
|---|---|
| AF | 87 |
| LF | 94 |
| TF | 96 |
| CFs (excluding central control system and common instrumentation) | 98 |
| Central control system and common instrumentation | 98 |
| Total (product) | 75 |

Taking into account the scheduled maintenance time, the operational availability requirement of 70% over the calendar year is translated into the inherent availability requirement of 75% over the 11 months of scheduled operation time. This inherent availability requirement, allocated among the facilities, can be seen in table 4.

The compliance with these inherent availability requirements together with the respect of the time allocated for the scheduled maintenance tasks would guarantee the plant availability requirement.

Generally RH is adopted to prevent radiological exposure of personnel during the maintenance operations as well as during experimental activities [101]. When the radiation field is above the hands-on limit of the radiation protection guidelines of the ICRP60 (i.e. >10 $\mu$Sv $h^{-1}$), different approaches can be used in the function of the dose rates expected in each area: the use of RH technologies, the use of local shielding, the maintenance performance to be carried out by the worker in shifts or delayed until the dose decreases sufficiently. Some IFMIF components require regular and scheduled maintenance, such as the annual long shut down, as well as replacement in case of failure.

The classification of components, from the RH point of view, is a rather complex activity since it depends on a number of factors like the RAMI analyses, the dose rate maps and zoning of the area where the components are installed. To help in the definition of maintenance priorities IFMIF has adopted the maintenance classification of components set up for the ITER project as follows: (1) RH class first for components requiring regular planned replacement; (2) RH class second for components that are likely to require repair or replacement, (3) RH class third for components that are not expected to require maintenance or replacement during the lifetime of the facility but would need to be replaced remotely should they fail; and (4) RH class fourth for components that do not require RH.

A basic assumption of the defined maintenance strategy is that the specimens irradiated will not be required to be immediately re-irradiated and included in an ensuing irradiation campaign. Almost all the RH maintenance activities for the IFMIF components will be performed in the AC and in the TMHC [109, 110] (see figure 24).

Efficient RH equipment in the TF is essential to meet the expected availability. The RH system for TF is installed in the AC and the TMHCs to perform operations on the irradiated TA and TMs. RH devices and robotics are needed inside the AC. The thickness of the walls all around the AC will be 1.5 m of concrete as BS. During the irradiation phase, the AC is accessible for operators; during the maintenance periods, the AC is accessible only if the TC and the TMHCs are closed and it is empty of activated components. Inert atmosphere inside the AC is not required. Nevertheless, a covered duct is designed in the AC floor to guide the gas and power supply and signal lines from the TC to the surrounding utility rooms. The AC has lead glass windows for direct visual access to the operations inside the AC. The current dimensions of the AC are 62 m × 9 m × 13 m.

Under the validation activities, RH trials of the TA have been in place with the construction of a mock-up to simulate the remote replacement of a removable backwall plate based on the bayonet concept and detachment and reconnection of flanges fastened with a dedicated connection system [85]; in turn, a dedicated orbital laser welding machine has been developed to remotely weld and cut a lipseal flange. The process has been validated using a mock-up of the IFMIF inlet pipe.

## 7. IFMIF construction schedule and cost

The construction schedule for the IFMIF/CODA (construction, operation and decommissioning activities) phase, set out below, (see figure 25) has been developed. In particular, lessons learnt from the actual design, construction and commissioning activities, being carried out as part of the EVA phase, have been factored into the planning to establish a solid time basis for the estimate.

The actual plan will strongly depend on the organisation and arrangements that will be put in place for the design, procurement, construction and commissioning, as well as on the licensing procedure. Future decisions reached by the parties involved may confirm or alter the assumptions that have led to the present proposed schedule.

The construction and licensing framework is defined by the regulations in force in the country where IFMIF will be built, commissioned and operated. The construction can only start when the license is issued by the country regulatory authority. It is thus anticipated that site selection is a critical milestone in the CODA schedule (it is assumed that the site shall be selected no later than one year after CODA initiation).

The overall schedule shown before represents a reference scenario, which is a success-oriented schedule of design, procurement, construction, assembly and commissioning of IFMIF. It is concluded that the construction period that leads up to the operation of the two parallel accelerators at full current (2 × 125 mA) is ten years from the official initiation of the CODA phase.

The IFMIF/CODA cost estimate was jointly prepared by the different contributing countries and supported by industry collaboration based upon a specific division of work packages and costing assumptions. A summary of the cost outcomes resulting from Monte Carlo simulations (MCSs) conducted during this IFMIF/EVEDA phase are presented hereafter. A total of around 1000 cost elements were estimated and the results converged by better than 0.05% after more than 3000 iterations.

The IFMIF construction cost up to the operation phase is estimated to be 1062 MICF, while the annual operation cost is set to be around 99 MICF (see table 5). The IFMIF conversion factor (ICF) is used to provide a uniform cost summary given by each party (reporting cost in their local currency). The unit 1 ICF corresponds to 1 Euro = 120 Yen as of January 2013.

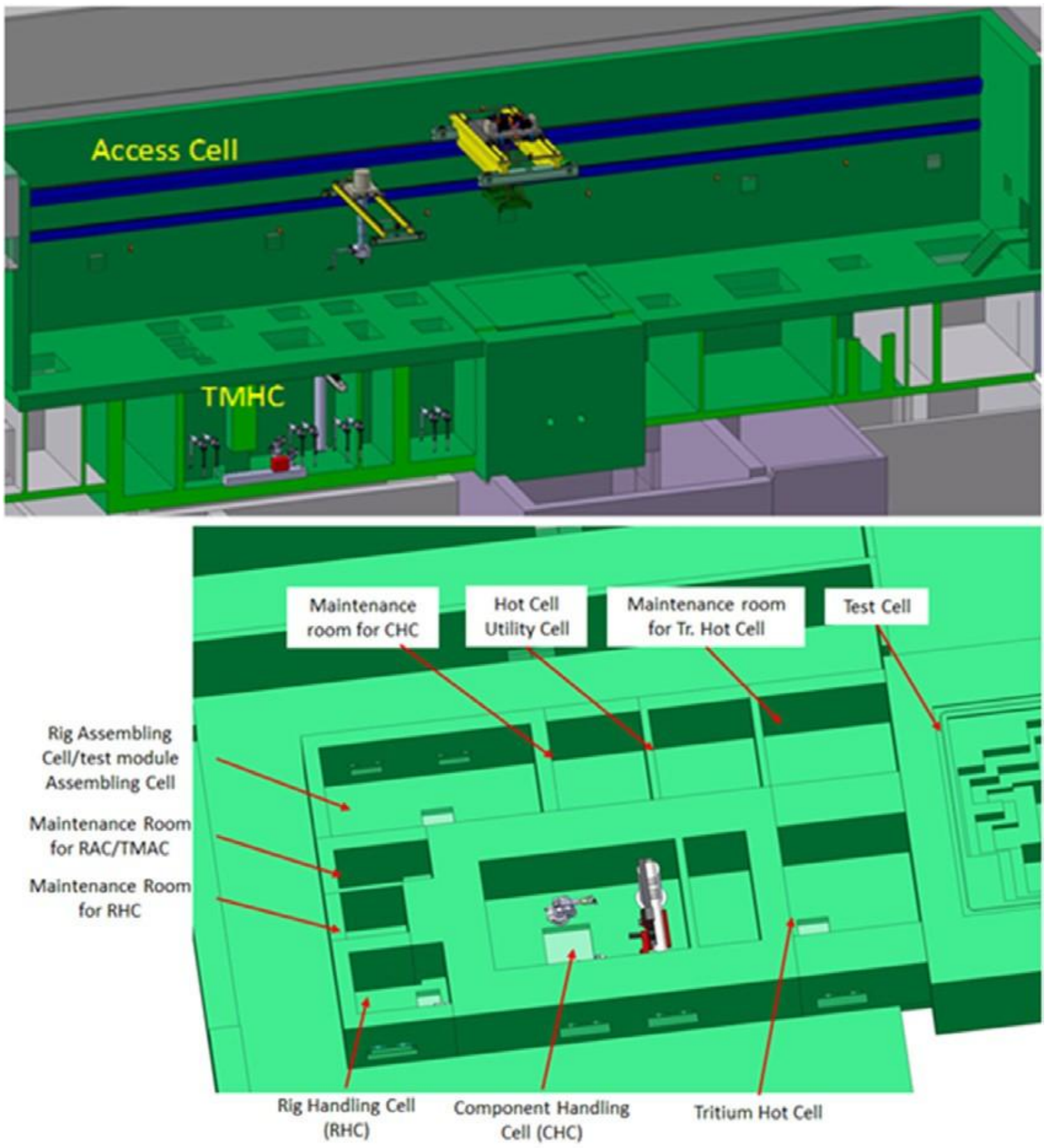

**Figure 24.** Top: overview of the AC and the TMHCs with the RH equipment installed. Left: overview of the TMHCs.

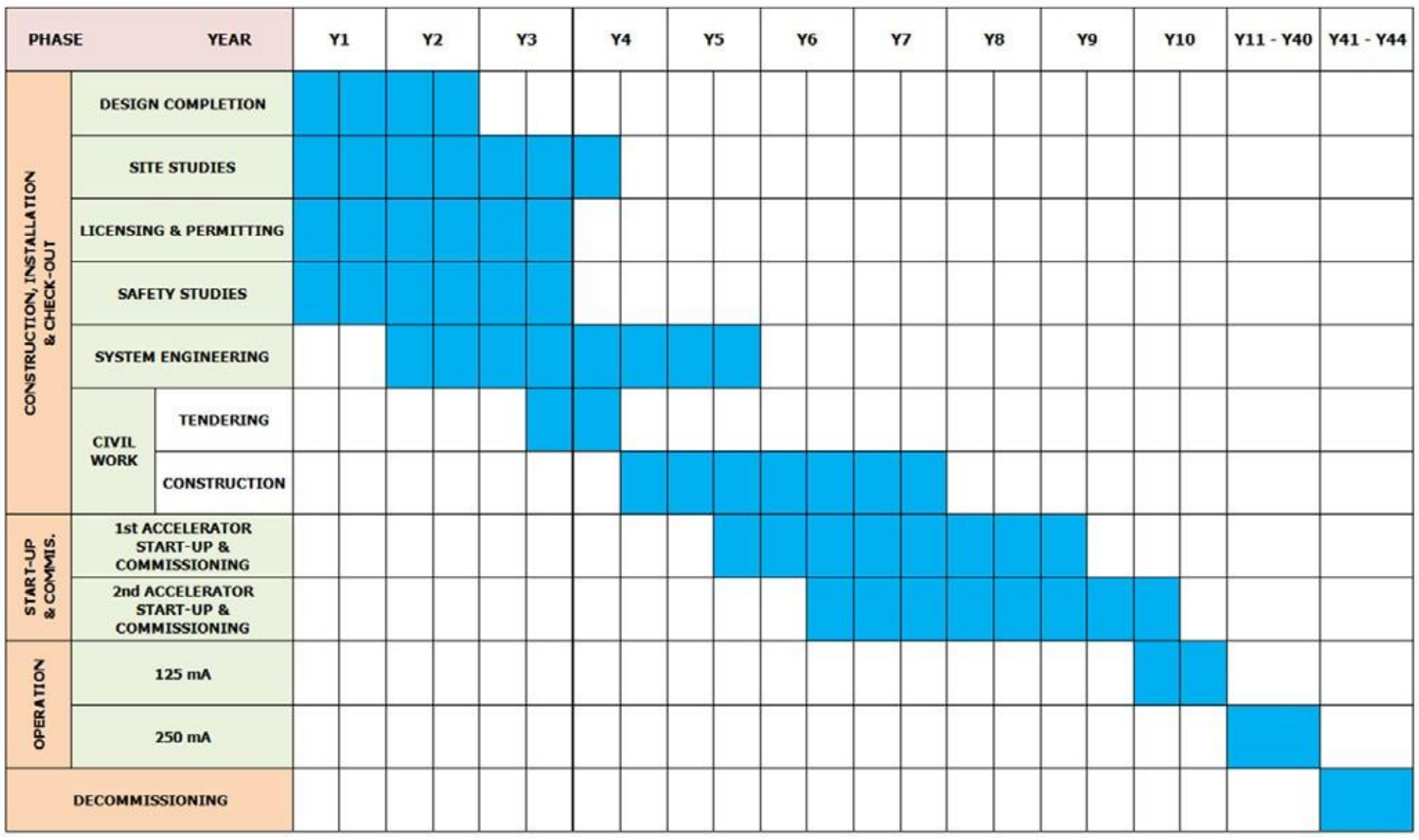

**Figure 25.** IFMIF CODA top level schedule.

The annual cost profile from the beginning of the construction phase to the 250 mA operation (until the end of Y12) is depicted in figure 26.

It is to be noted that the present cost estimation is considered to be precise since it takes into account the real cost of constructed prototypes, and the cost estimation for IFMIF has consistently reached similar values in all former project phases. The estimated cost is in the same range for equivalent facilities like the 1.4 billion US dollars reported by the SNS [111], which is presently under operation, or the 1.8 billion Euros cost estimation of the European spallation source [112] presently under construction.

**Table 5.** MCS Results (no escalation) for the IFMIF/CODA Phases.

| | MCS Cost (MICF) - no escalation | | |
|---|---|---|---|
| Cost Element | Mean value | P50 | P80 |
| IFMIF Construction, Installation & Check-out Phase | 958 | 951 | 1039 |
| IFMIF Start-Up & Commissioning phase | 104 | 103 | 113 |
| IFMIF Operation phase | 2977 | 2955 | 3236 |
| IFMIF Decommissioning phase | 154 | 153 | 171 |

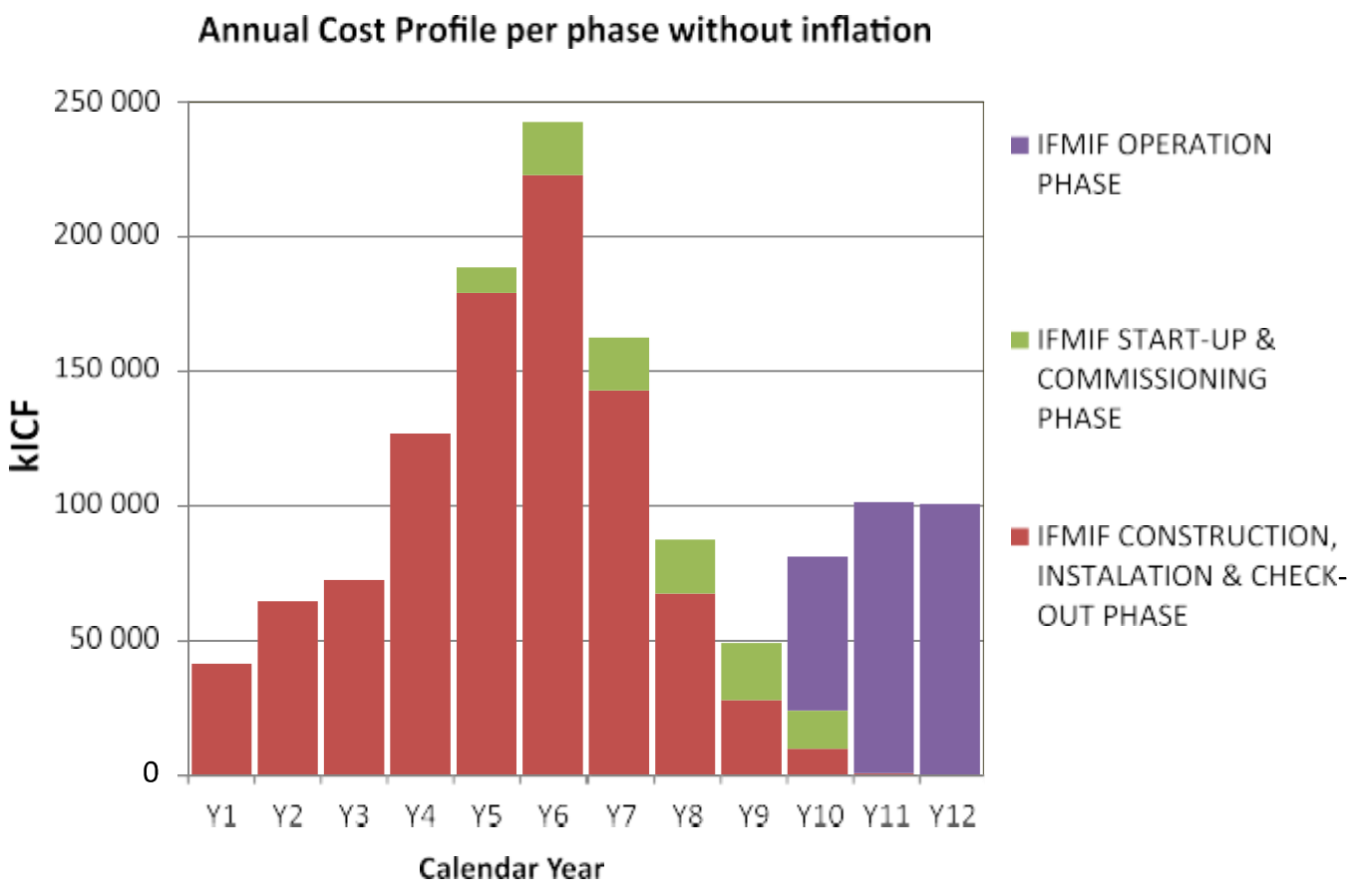

**Figure 26.** Annual cost profile for the construction phase through the operation phase.

## 8. Conclusions

A design for the IFMIF is made available accomplishing the EDA phase under IFMIF/EVEDA. In the ensuing construction phase, additional engineering efforts will be required to exploit the pending results in the still progressing EVA phase within the framework of the Broader Approach agreement and to adapt the present proposed design to the specificities of the construction site.

The success of the EDA phase of IFMIF, accomplished on schedule within the six years allocated, with a design backed by the parallel successful validation activities of the main technological challenges within the on-going EVA [19], allows us to reliably anticipate its construction in terms of schedule and costs.

In IFMIF's safety approach, the priority has been assigned to prevent accidents by a robust design to control hazards and to mitigate consequences of postulated initiating events (the safety principle 'defence in depth' has been integrated during this EDA phase). The safety analysis has been performed individually to each of the IFMIF's facility, with the exception of the PIEF (interfaces at the Plant level of the latter are to be studied in future phases). The PDD document is supported by 35 different DDDs of all the sub-systems. In addition, a full 3D modelisation and thorough product breakdown structure of the full facility has also been developed. A careful 'risk register', including potential risks in this on-going EVA phase has also been prepared.

The wealth of the constituent reports, listed in figure 3 will allow us to enter into the construction phase in a smooth manner. The CDR [18], exploiting the design mastery established throughout some three decades of continuous R&D activities, served as sound ground for this definitive EDA phase on the way to construction. Nevertheless, in the EDA phase significant advancements have been introduced into the design of the sub-systems of the five major systems, resolving technical issues remaining from previous design phases [24], such as: (1) the irradiation modules no longer have a shielding function, and as a result, the irradiation has significantly gained in flexibility providing greater ease in module positioning; (2) the RH equipment has been improved allowing an increase in reliability and a decrease in cost; (3) the DTL in the AF has been replaced by a superconducting radio-frequency linac, with a significant reduction in beam losses and operational costs; as a consequence, the RF system could be better modularised; (4) the QT of the lithium loop, previously included inside the TC, has been re-located outside, with a reduction of the tritium production and, in addition, the operations required to exchange the QT, in case of failure, have been simplified; (5) the lithium loop now has two intermediate secondary cooling oil circuits, reducing the risks associated with the presence of lithium, reducing the thermal gradients in the heat removal system and avoiding potential water boiling in case of loss of water flow; (6) the liner and BS of the TC can now be cooled with water, enhancing the efficiency and economy of the related sub-systems by adding a liner in the lithium loop room; (7) the lithium loop now has a by-pass, allowing more flexibility during its operation; most of the safety critical operations linked to the manipulation of the irradiated modules and TA have been concentrated in a relatively small hot cell; (8) the injector design has been improved by adding a supplementary extraction electrode gaining in availability and a chopper that will ease the commissioning of the accelerators; and (9) the maintenance strategy has been modified to allow a shorter yearly stop of the irradiation operations and a more careful management of the irradiated samples.

All world fusion power roadmaps expect to have MWs in the electrical grid in the 40s of this century; for this to happen, decades old remaining open questions on structural materials behaviour when facing DT reactions are to be unravelled next decade; otherwise the timely accomplishment of the design of any next step after ITER is not possible.

The combination of the following three factors: (1) the second order cost of a Li(d,xn) neutron source compared with the cost of a fusion reactor; (2) the maturity of the design of IFMIF thanks to the successful achievements of the IFMIF/EVEDA phase; and (3) the indispensability of understanding the behaviour of plasma facing components under 14 MeV neutrons, should result in the timely construction of a Li(d,xn) fusion relevant neutron source adapted to the fusion community needs.